%% file: 00_main.tex
\documentclass[manuscript, screen, sigconf]{acmart}
\AtBeginDocument{%
  \providecommand\BibTeX{{%
    \normalfont B\kern-0.5em{\scshape i\kern-0.25em b}\kern-0.8em\TeX}}}

\setcopyright{acmcopyright}
\copyrightyear{2018}
\acmYear{2018}
\acmDOI{XXXXXXX.XXXXXXX}

\acmConference[Conference acronym 'XX]{Make sure to enter the correct
  conference title from your rights confirmation emai}{June 03--05,
  2018}{Woodstock, NY}
\acmBooktitle{Woodstock '18: ACM Symposium on Neural Gaze Detection,
 June 03--05, 2018, Woodstock, NY} 
\acmPrice{15.00}
\acmISBN{978-1-4503-XXXX-X/18/06}

\usepackage{xcolor}
\usepackage{multirow}
\usepackage{array}

\usepackage{xspace}
\newcommand{\system}{SimStep\xspace} 
\newcommand{\systems}{SimStep's\xspace}

\definecolor{skyblue}{RGB}{223, 239, 245}

\usepackage{comment}
\usepackage{soul}
\usepackage{pifont}
\usepackage{svg}
\usepackage{listings}
\usepackage{listingsutf8}
\usepackage{mdframed}
\usepackage{setspace}
\definecolor{lightbrown}{RGB}{255,249,241}
\newenvironment{promptbox}
    {\begin{mdframed}[backgroundcolor=lightbrown,linecolor=white]
        \begin{small}
            \begin{spacing}{1}}
    {\end{spacing}
    \end{small}
    \end{mdframed}
    }

    \lstdefinelanguage{json}{
  basicstyle=\ttfamily\small,
  showstringspaces=false,
  breaklines=true,
  frame=none,
  backgroundcolor=\color{white},
  literate=
   *{0}{{\color{blue}0}}{1}
    {1}{{\color{blue}1}}{1}
    {2}{{\color{blue}2}}{1}
    {3}{{\color{blue}3}}{1}
    {4}{{\color{blue}4}}{1}
    {5}{{\color{blue}5}}{1}
    {6}{{\color{blue}6}}{1}
    {7}{{\color{blue}7}}{1}
    {8}{{\color{blue}8}}{1}
    {9}{{\color{blue}9}}{1}
    {:}{{\color{black}:}}{1}
    {,}{{\color{black},}}{1}
    {"}{{\color{red}"}}{1},
}

\lstdefinestyle{compactjson}{
  language=json,
  basicstyle=\ttfamily\small,
  aboveskip=0pt,
  belowskip=0pt,
  xleftmargin=1em,
  numbers=none
}
\begin{document}


\title[Chain-of-Abstractions for Authoring AI Simulations]{SimStep: Chain-of-Abstractions for Incremental Specification and Debugging of AI-Generated Interactive Simulations}

\author{Zoe Kaputa}
 \affiliation{%
  \institution{Stanford University}
   \country{USA}
 }
 \email{kaputa@stanford.edu}

 \author{Anika Rajaram}
 \affiliation{%
  \institution{The Harker School}
   \country{USA}
 }
 \email{26anikar@gmail.com}

 \author{Vryan Almanon Feliciano}
 \affiliation{%
  \institution{Stanford University}
   \country{USA}
 }
 \email{vgfelica@stanford.edu}

\author{Zhuoyue Lyu}
 \affiliation{%
  \institution{University of Cambridge}
   \country{UK}
 }
 \email{zl536@cam.ac.uk}

 \author{Maneesh Agrawala}
 \affiliation{%
  \institution{Stanford University}
   \country{USA}
 }
 \email{maneesh@cs.stanford.edu}

\author{Hari Subramonyam}
 \affiliation{%
  \institution{Stanford University}
   \country{USA}
 }
 \email{harihars@stanford.edu}

\renewcommand{\shortauthors}{Kaputa et al.}

\begin{abstract}
Programming-by-prompting with generative AI offers a new paradigm for end-user programming, shifting the focus from syntactic fluency to semantic intent. This shift holds particular promise for non-programmers such as educators, who can describe instructional goals in natural language to generate interactive learning content. Yet in bypassing direct code authoring, many of programming's core affordances---such as traceability, stepwise refinement, and behavioral testing---are lost. We propose the \textit{Chain-of-Abstractions (CoA)} framework as a way to recover these affordances while preserving the expressive flexibility of natural language. CoA decomposes the synthesis process into a sequence of cognitively meaningful, task-aligned representations that function as checkpoints for specification, inspection, and refinement. We instantiate this approach in \system, an authoring environment for teachers that scaffolds simulation creation through four intermediate abstractions: Concept Graph, Scenario Graph, Learning Goal Graph, and UI Interaction Graph. To address ambiguities and misalignments, \system includes an inverse correction process that surfaces in-filled model assumptions and enables targeted revision without requiring users to manipulate code. Evaluations with educators show that CoA enables greater authoring control and interpretability in programming-by-prompting workflows.
\end{abstract}



% \keywords{datasets, neural networks, gaze detection, text tagging}

\begin{teaserfigure}
  \includegraphics[width=\textwidth]{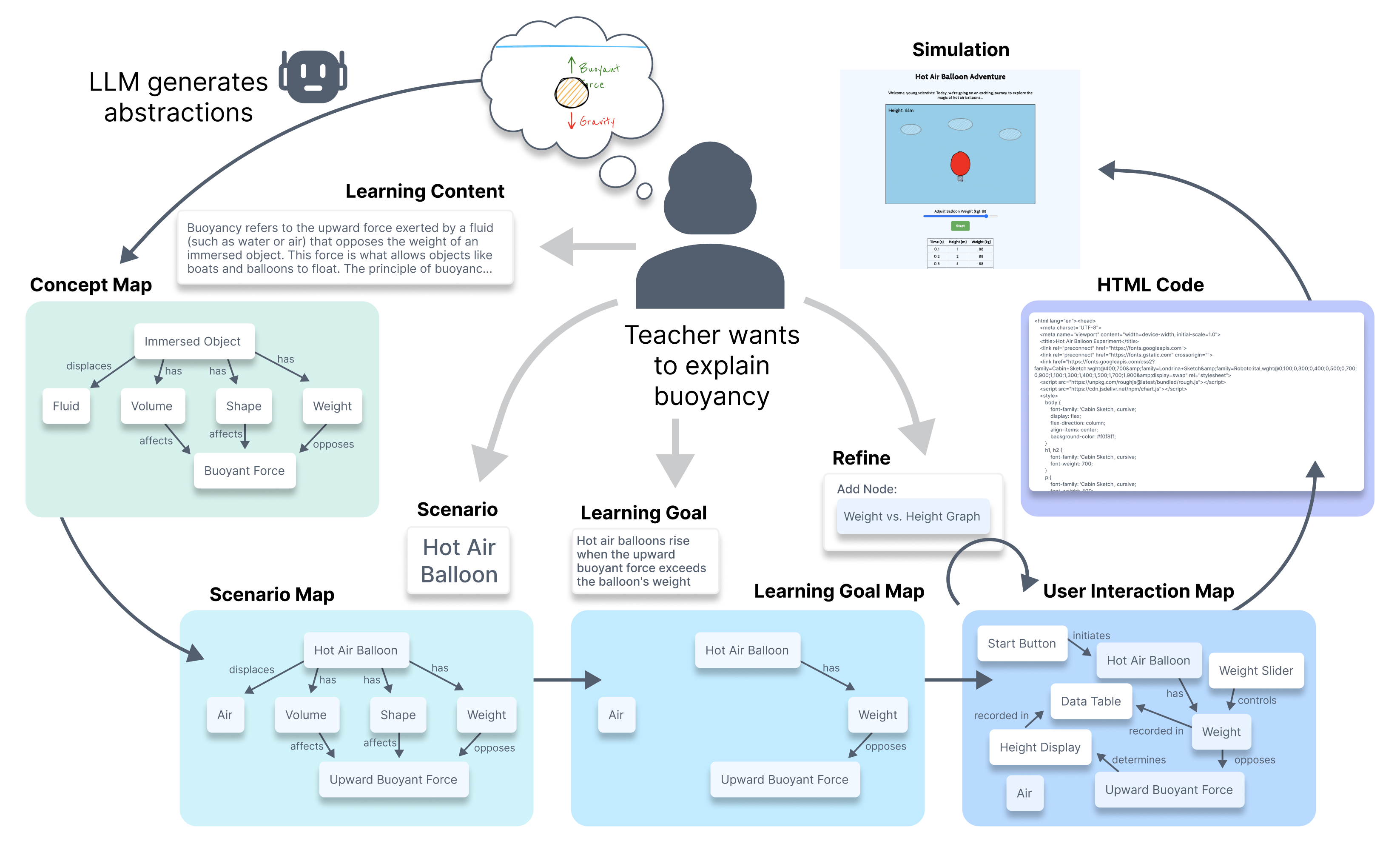}
  \caption{A teacher can use \system to generate interactive simulations that are accurate, integrate students' context, and reflect straightforward learning goals.}
  \Description{A diagram showing the overall flow of a teacher generating a simulation from their initial idea. The flow is represented by a circle of abstractions that surround the teacher. These representations include the Concept Graph, the Scenario Graph, the Learning Goal Graph, and the User Interaction Graph. This circle ends with the HTML Code and the Simulation itself. The user is connected to various abstractions through arrows indicating user choices.}
  \label{fig:teaser}
\end{teaserfigure}


\maketitle

\input{01_intro}
\input{02_framework}
\input{03_ux}
\input{04_system}
\input{05_user_evaluation}
\input{06_techincal_evaluation}
\input{07_discussion}
\input{08_relatedwork}
\input{09_conclusion}

\bibliographystyle{ACM-Reference-Format}
\bibliography{99_refs}

\input{10_appendix}

\clearpage
\appendix

\end{document}

%% file: 01_intro.tex
\section{Introduction}

\textit{Programming-by-prompting} with generative AI promises to democratize code creation~\cite{Jiang2022DiscoveringTSA}. It shifts the focus from writing \textit{syntactically} correct instructions to expressing \textit{semantically} meaningful intent~\cite{song2023empirical}. For non-technical experts such as educators, this approach is potentially transformative. Rather than learning formal programming constructs, teachers can describe goals, concepts, and exemplar scenarios in natural language and create rich interactive learning experiences for their students~\cite{barnett1968natural}. For instance, a teacher might prompt \textit{``give me a simulation to teach about buoyancy by showing hot-air balloons rise and fall,''} and receive a functional interactive simulation aligned with their lesson (see Figure~\ref{fig:teaser}). Here the authoring interface is natural language itself allowing teachers to express program specifications using the same language and vocabulary they use to plan lessons, explain ideas, and promote active student engagement. 

However, in abstracting away the syntactic interface for natural language, something essential is lost. Traditional syntactic programming interfaces have certain affordances. Every behavior must be explicity specified, requiring \textit{users} to resolve ambiguities, consider conditionals and alternatives, and translate high level goals into concrete operational logic. It reveals gaps in understanding and makes assumptions visible. Prompting, by contrast, allows ideas to remain \textit{diffuse} and critical details are often left unstated~\cite{subramonyam2024bridging}. The generative AI model must infer what is missing and resolve underspecifications, and those inferences may not always be aligned with goals and intents. In the previous example, the prompt about buoyancy simulation might yield a visually appealing animation, yet omit essential causal relationships, misrepresent scientific principles, or fail to support student interactions that deepen understanding. 

Further, regardless of the interface, certain core programming tasks---such as testing behavior, debugging errors, and refining algorithms---remain. Syntactic interfaces afford traceability and control by exposing the program's structure through elements such as function definitions, control flow, and variable states, enabling the programmer to trace and debug the behavior step-by-step. In the buoyancy simulation, the programmer can inspect the exact buoyancy equation governing the motion of the hot-air balloon or modify conditional logic to assess edge cases. In contrast, natural language \textit{collapses} these structures into textual abstractions, making it difficult to isolate errors, reason about behavior, or understand how changes to the prompt affect the generated code. 

Existing approaches to alleviate these challenges largely cluster around three main strategies, each targeting different aspects of the prompt-to-code process. First, \textit{prompt engineering} techniques focus on improving the prompt quality through carefully worded instructions, structured templates, or few-shot examples that guide and constrain the model's interpretation~\cite{white2023prompt}. While often helpful, prompt engineering places a cognitive burden to envision how the model ``thinks,'' which can be especially challenging for non-technical users such as teachers~\cite{subramonyam2024bridging, tan2024more}. Second, \textit{post-generation debugging} approaches allow users to refine code either by mapping back the code to the prompt~\cite{Liu2023WhatIW}, providing dynamic interfaces for editing~\cite{angert2023spellburst}, or directly inspecting and editing the generated code to correct errors~\cite{tan2023copilot}. A third strategy, reasoning trace externalization, includes techniques such as \textit{chain-of-thought}~\cite{wei2022chain} and \textit{chain-of-verification}~\cite{dhuliawala2023chain}, which aim to improve model performance and interpretability by generating intermediate steps in natural language. While often described as revealing the model’s ``reasoning,'' these techniques do not expose internal symbolic computations or structured deliberation, but rather simulate step-by-step patterns that improve alignment with desired outputs~\cite{lyu2023faithful}. As such, they operate within linguistic scaffolding, offering no direct support for users to specify or manipulate the underlying system behavior.

As an alternative, we propose that programming-by-prompting should not solely rely on linguistic interactions and abstractions. But instead, it should be supported by \textbf{task-level representational abstractions} that externalize and structure user intent. This approach is grounded in the theory of distributed cognition, which holds that complex tasks are accomplished not by internal reasoning alone, but through interactions with external representations that guide and transform cognitive work~\cite{hutchins1995cognition}. In our context, these task abstractions (1) align with natural task decomposition, and (2) act as cognitive checkpoints---distinct stages where users can interpret, manipulate, and refine system behavior.  The checkpoints allow users to verify and refine behavior, test assumptions, and refine outcomes, all without requiring formal programming. 

In this paper, we introduce \system, a system that operationalizes this approach through a chain-of-abstractions (CoA) framework. Grounded in the task of authoring educational simulations, \system guides educators through a structured sequence of \textit{domain-aligned} representations---including a Concept Graph, Scenario Graph, Learning Goal Graph, and UI Interaction Graph---each designed to surface and refine different dimension of the intent (Figure~\ref{fig:architecture}). These representations allow users to incrementally clarify and operationalize their ideas, while also providing structure for the model to reason from. An \textit{underspecification resolution engine} further supports this process by identifying ambiguous or missing elements, enabling users to \textit{inspect} model assumptions, \textit{test} behavior in a \textit{guided} manner, and iteratively \textit{steer} the generation toward their instructional goals. At every step, \system keeps the human in the loop, as an active participant in shaping interpretations, validating assumptions, and steering the generation. Our approach recovers key affordances of traditional programming such as traceability, testability, and control through accessible, task-specific abstractions.

Our key contributions are: (1) a conceptual framework that repositions programming-by-prompting as authoring through task-level abstractions, emphasizing the role of human-in-the-loop reasoning and correction; (2) the design and implementation of \system, an instantiation of this framework that enables educators to incrementally specify, test, and revise AI-generated simulations without writing code; and (3) technical and empirical evaluation demonstrating how \system supports pedagogical alignment, reduces authoring complexity, and provides control in content creation.

%% file: 02_framework.tex
\section{Conceptual Framework for CoA}\label{sec:framework}

To ground prompt-to-code authoring as a human-in-the-loop process, we formalize our approach using a distributed cognition lens~\cite{hutchins1995cognition}. Distributed cognition theory views cognitive processes as extending beyond individual minds, operating across people, artifacts, and representational media. In Hutchins' study of ship navigation, for example, navigational charts, instruments, and logs serve not merely as information displays, but as cognitive artifacts that structure and distribute reasoning over time and across individuals~\cite{hutchins1995cognition}. 

Our Chain-of-Abstractions (CoA) framework adopts this perspective by treating intermediate representations not simply as steps in a pipeline but as structured cognitive checkpoints where reasoning is transformed and shared between humans and AI. Each abstraction in the CoA supports distinct types of cognitive work---such as articulating domain knowledge, defining causal structure, or specifying interface logic---and affords both interpretability for the human and tractability for the model. What qualifies this decomposition as a form of distributed cognition is not merely that the process is staged, but that each stage enables coordination between agents (human and AI), externalizes internal thought processes, and provides a representational substrate for validation, revision, and semantic control. In this sense, CoA reflects a system of representations that scaffolds joint reasoning across agents, aligning with core principles of distributed cognitive systems.

\subsection{Human Guided Forward Transformations}
Let $P$ denote a user's initial natural language prompt and $C$ the final executable code. Between them lies a sequence of representational abstractions $\mathcal{A} = \{A_1, A_2, ..., A_n\}$, each representing a task checkpoint where intent is transformed, clarified, and refined. In prior work, several key types of abstractions have been proposed including concept graphs, scene graphs, task decomposition structures, etc.~\cite{Weyssow2022BetterMT,Zhu2020HierarchicalPF,Armeni20193DSG,Ritschel2022CanGD}. These abstractions are derived through transformations $T = \{T_1, ..., T_{n+1}\}$, and we propose they involve a collaboration between human agents ($H$) and machine agents ($M$) powered by large language models. We formalize the collaborative transformation process as:
\[
P \xrightarrow{T_1^{(M)}} A_1 \xrightarrow{\text{Refine}^{(H)}} A_1' \xrightarrow{T_2^{(M)}} A_2 \xrightarrow{\text{Direct}^{(H)}} A_2' \xrightarrow{\cdots} C
\]

Each abstraction $A_i$ serves as a \textit{task checkpoint} that exposes the model's current understanding of the task. Functionally, each abstractions is characterized by its \textit{visibility}---how observable and legible its structure is to the user, its \textit{manipulability}---how easily the user can adjust or modify its components, and its \textit{fidelity}---how accurately it encodes the user’s goals or task-relevant system logic. These properties collectively determine the abstraction's  effectiveness in supporting meaningful human oversight and intervention within the generative workflow. 

Further, we define four core \emph{forward operations} that the human agent $H$ performs at each checkpoint $A_i$:

\vspace{1mm}
\begin{itemize}
    \item \textbf{Inspect}: Examine the abstraction $A_i$ to understand what the model has inferred, identify mismatches with intent, and diagnose possible errors or oversights.
    \item \textbf{Refine}: Modify $A_i$ to produce $A_i'$, correcting assumptions, adding missing components, adjusting relationships, or removing irrelevant elements.
    \item \textbf{Validate}: Assess whether $A_i'$ meets instructional or behavioral goals, ensuring semantic coherence, completeness, and alignment with domain expectations. 
    \item \textbf{Direct}: Provide guidance for the next transformation $T_{i+1}^{(M)}$ by specifying priorities, constraints, or features that must be preserved in downstream abstractions. 
\end{itemize}
\vspace{-1mm}

\subsection{Inverse Correction}
The forward transformation pipeline produces a sequence of intermediate abstractions from a natural language prompt. Each abstraction $A_i$ represents a progressively refined interpretation of user intent, narrowing ambiguity and structuring information for downstream synthesis. Ideally, underspecified details would be surfaced and resolved at appropriate points along this chain. However, in practice, generative models often defer filling in those details until the final stages, particularly at the code level where the model must commit to specific values, logic, or visual behavior. This can lead to implicit assumptions being introduced without the user's awareness, as those decisions were not made explicit in earlier abstractions.

While introducing intermediate abstractions helps reduce ambiguity and support human-in-the-loop correction, it also introduces a design tradeoff. Increasing the number of abstractions, or lowering their level of specificity, can overwhelm users with too many representational layers or expose them to implementation-level detail. For instance, block-based environments like Scratch~\cite{maloney2010scratch} provide visual structure but often require users to reason about program flow, variable state, and low-level operations. This reintroduces syntactic and procedural complexity, defeating the purpose of semantic-level prompting. In contrast, our approach favors task-level abstractions that align with how users naturally organize their thinking, aiming for a balance between expressive control and cognitive accessibility. 

However, to recover and revise these hidden assumptions, we introduce a complementary \textbf{inverse process}, which includes a new set of \textit{targeted} abstractions $\mathcal{B} = \{B_1, B_2, ..., B_m\}$. The same abstraction $X$ may exist in both $\mathcal{A}$ and $\mathcal{B}$, but these sets can also include unique abstractions. The inverse correction process begins by realizing the $B_i$ that is most suitable for correcting the system's assumption. Then, the user can \textbf{Inspect}, \textbf{Refine}, and \textbf{Validate} $B_i$ as described for abstractions in $\mathcal{A}$. After producing $B_i'$, a direct translation $T^{(M)}$ exists to transform $B_i'$ into $C'$.
\[
C \xrightarrow{\text{Realize}} B_i \xrightarrow{\text{Refine}^{(H)}}
B_i' \xrightarrow{T^{(M)}} C'
\]

This revision occurs at the abstraction level rather than the code level, allowing users to correct high-level misunderstandings without needing to inspect low-level syntax. Let $\Omega(X)$ denote the set of all valid implementations (e.g., simulation code) that are compatible with a representation $X$, and let $U(X)$ represent the degree of underspecification in $X$. If a representation is vague or abstract, $\Omega(X)$ will be large, indicating that many different implementations are possible, and $U(X)$ will be high, reflecting the ambiguity of the representation.

As the synthesis pipeline progresses from the natural language prompt $P$ to the final code $C$, each transformation incrementally reduces both the ambiguity and the number of compatible implementations:
\[
\Omega(P) \supset \Omega(B_1) \supset \Omega(B_1') \supset \cdots \supset \Omega(C)
\]
\[
U(P) > U(B_1) > U(B_1') > \cdots > U(C)
\]

This formalism reflects how intent becomes increasingly concrete. Note that this process is not intended to surface all hidden details but to expose and resolve only those underspecifications that result in incorrect behavior. It turns abstractions into bi-directional interfaces that can be used for forward synthesis as well as targeted recovery. 

%% file: 03_ux.tex
\section{User Experience}\label{sec:ux}

\begin{figure*}
  \centering
  \includegraphics[width=0.85\linewidth]{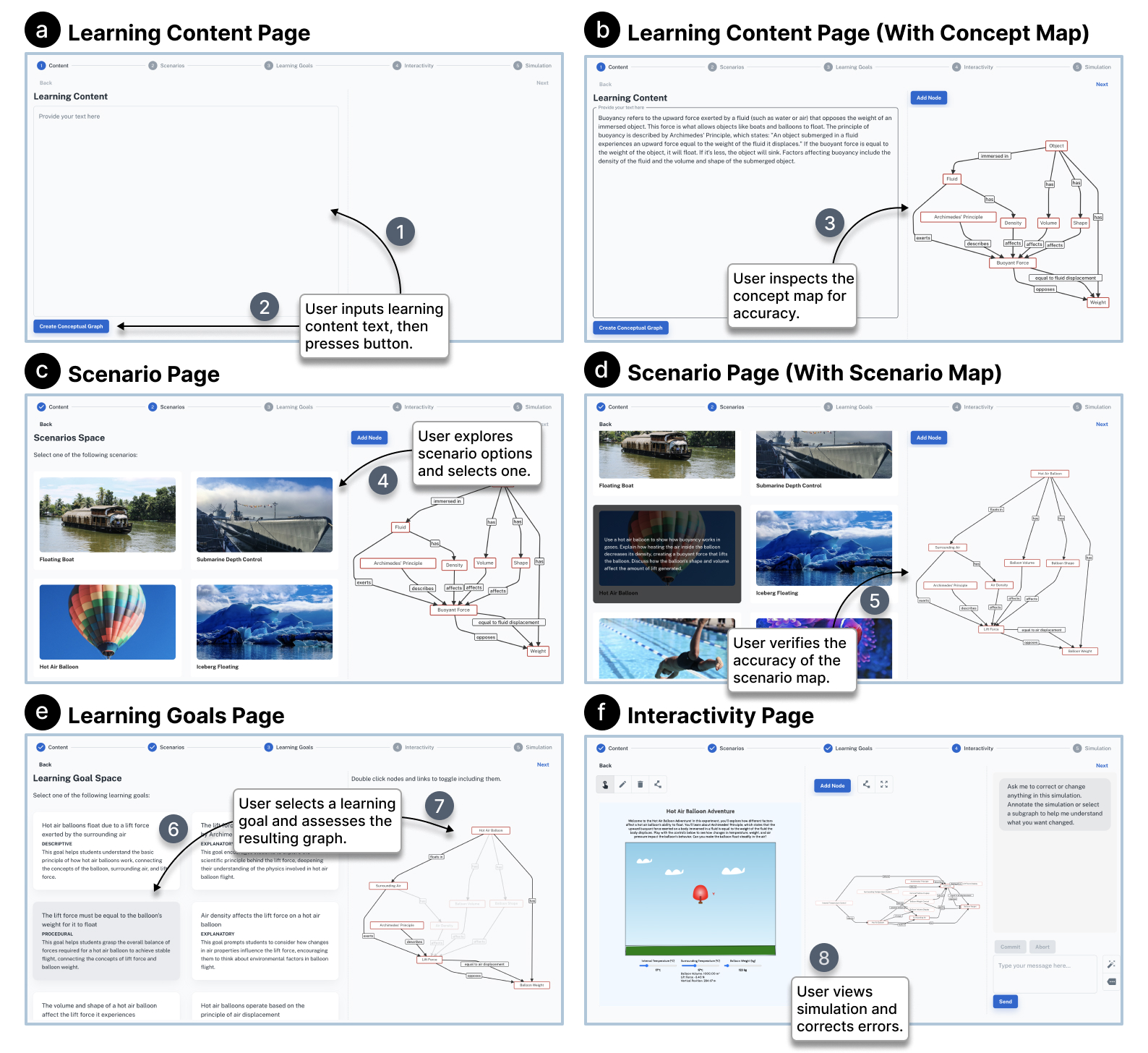}
  \caption{The main screens of the \system application include a Learning Content Page (a and b), a Scenario Page (c and d), a Learning Goals Page (e), and an Interactivity Page (f).}
  \Description{A grid containing 6 screenshots of the \system application, with 3 rows and 2 columns. }
  \label{fig:screens}
\end{figure*}

To realize the CoA framework in Section~\ref{sec:framework}, we developed \system, an tool that allows educators to author interactive simulations through a human-guided, step-by-step approach. When designing a simulation, a human expert typically approaches this task in stages: (1) identifying the core scientific concepts to teach, (2) searching for and mapping how they relate with potential experimental scenarios, (3) concretely defining the situated learning goals, and (4) considering the space of hypothesis to realize those goals and how learners will interact with the system. Based on this understanding, \systems interface follows a wizard-style design pattern using the four main stages in the task, providing appropriate representational abstractions in each stage. To explore the specific authoring and testing features, let us follow Mr. Carlos, a high school science teacher, as he creates a simulation using \system. 

\subsection{CoA Code Generation}
Mr. Carlos is working on his lesson plan to teach about \textit{buoyancy} and wishes to use an interactive simulation to promote active student engagement. Mr. Carlos opens \system on his web browser, which shows him the authoring interface with the \textit{text input step} open (Figure~\ref{fig:screens}a). This step has a text input box on the left and a collapsible panel to display the concept graph on the right. Using the prescribed science textbook for the course, Mr. Carlos copies the text about core principles of buoyancy and pastes it in the text box. Based on this text, \system automatically generates a \textbf{Concept Graph}---a visual model made up of nodes representing key concepts like Object's Weight, Buoyant Force $B$, Fluid Density $\rho$, and Displaced Volume $V$, connected by edges that encode relationships and equations (e.g., $m = \rho_{\text{object}} \times V$, $W = m \times g$). As Carlos \textbf{Inspects} the graph, he notices an issue: Buoyant Force is linked directly to Object's Mass, skipping the required relationship with Fluid Density and Displaced Volume. Using the graph editor, he deletes the incorrect link and adds two new nodes and connecting links to represent the correct equation (i.e., \textbf{Refine}): $B = \rho \times V \times g$. Once the concept graph accurately reflects the scientific model, Carlos proceeds to the next step.

\begin{figure*}
  \centering
  \includegraphics[width=\linewidth]{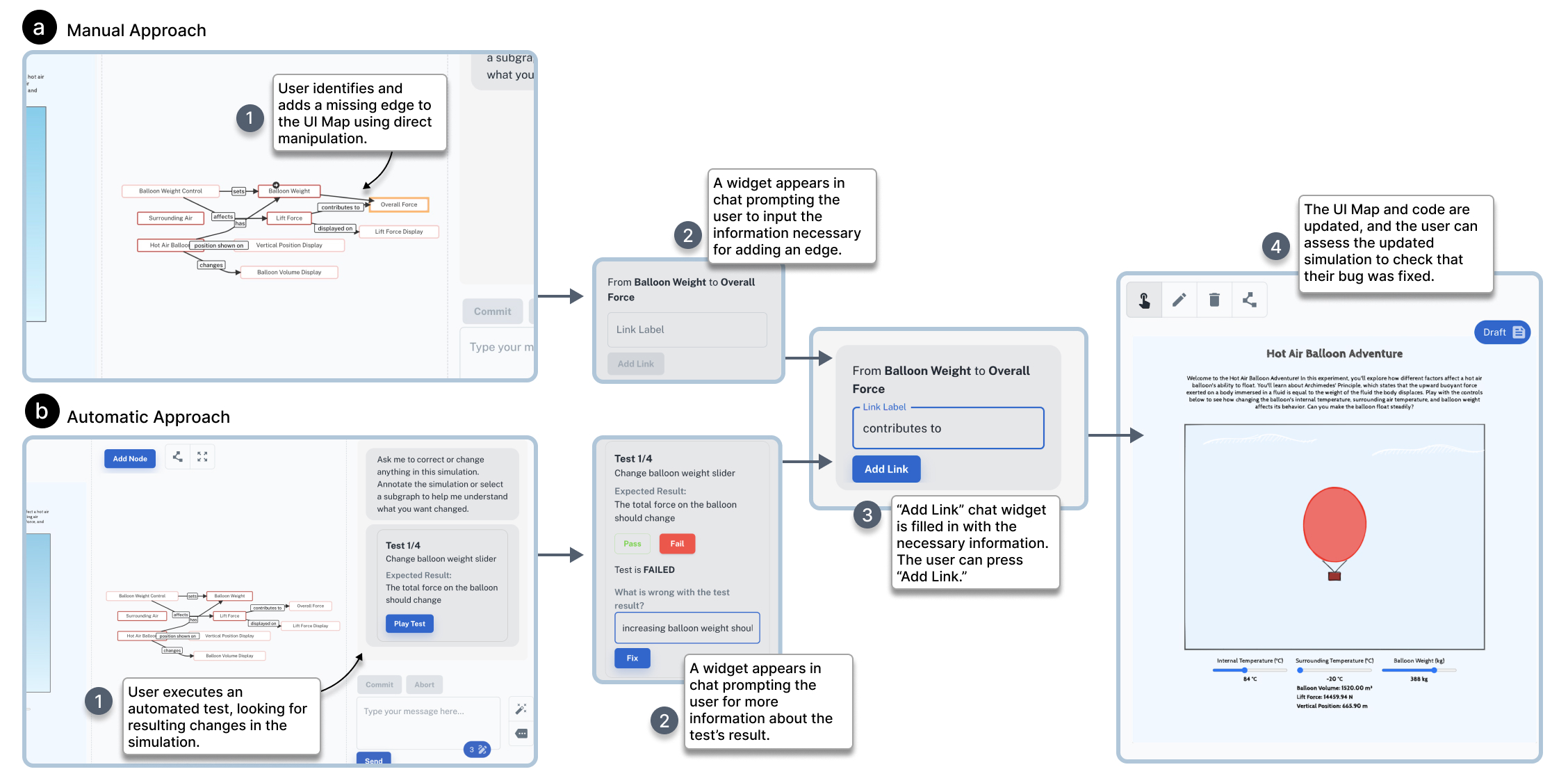}
  \caption{The debugging process can be split into two main approaches: the manual approach and the automatic approach, which provide the user varying levels of control and automation.}
  \Description{A diagram with two main portions: one labeled "Manual Approach" and one labeled "Automatic Approach." This diagram displays six screenshots from the app and explains through annotations that the user can use two approaches for correcting bugs; they can either directly update the abstraction that is incorrect using a chat widget, or they can tell the chat to fix an issue (perhaps discovered through automated test cases), which will pre-fill a widget for them. }
  \label{fig:widget-ux}
\end{figure*}

In the second step, Mr. Carlos is prompted to specify a desired experimental scenario (just like a human expert would do), which serves as the contextual foundation for generating the simulation code. To facilitate scenario grounding, \system uses the conceptual model to generates and displays a set of potential scenarios for teaching about buoyancy (Figure~\ref{fig:screens}c). Mr. Carlos notices that one of the scenarios is using \textit{hot air balloon}, and realizes that since his students recently saw hot air balloons rise at the annual city festival, it would be a great way to connect the abstract concept of buoyancy to something they had all experienced. Alternatively, Mr. Carlos can define his own scenario using the text input box, tailored to his understanding of his students. Once Mr. Carlos provides a scenario, system generates a \textbf{Scenario Graph}, a situation model representation (Figure~\ref{fig:screens}d) in which all the nodes and links in the conceptual model are instantiated with the context of hot air balloon. As before, Carlos \textbf{Inspects} the instantiated graph and proceeds to the learning goal selection by clicking the `Next' button. 

The learning goal selection panel offers Mr. Carlos a range of objectives tailored to the chosen scenario. For instance, \system might presents several objectives around (1) conceptual understanding focusing on key concepts and describing phenomenon (e.g., \textit{understand the relationship between air temperature inside the balloon and its buoyancy, and be able to describe how equilibrium is achieved}) , causal or explanatory understanding (e.g., \textit{explain how the decrease in air density inside the balloon leads to an increase in buoyancy, allowing the balloon to rise}), and procedural knowledge (e.g.,  \textit{effects of different heating methods on the rate of the balloon's ascent and procedural relationship between gradual heating and smooth altitude control}). On this panel (Figure~\ref{fig:screens}e), Mr. Carlos can select a learning goal to \textit{direct} next steps in the code generation, and as before explore the generated \textbf{Learning Goal Graph} which is derived from the situation model. From here, Mr. Carlos simply clicks the `Next' Button to see the final interactive simulation (Figure~\ref{fig:screens}f). To generate the simulation, \system uses the learning goals sub-graph and provides an \textbf{Interactivity Graph} including all of the controls and how controls link to behavior of elements in the situation model to meet the intended learning goals.

\subsection{Interactive Debugging and Refinement}

While the forward authoring path provides several affordances for aligning steering code generation, potential errors can still emerge in realizing the final interaction graph as simulation code, i.e., syntactic inference errors. The errors might include misaligned layout elements or mislabeled controls, errors where the model fills in unspecified details like slider ranges or animation timings incorrectly; or errors where UI elements fail to trigger the expected behaviors due to missing or flawed logic. To support testing, debugging and correcting these issues, \system offers several features including guided testing, and widget based error correction.

\subsubsection{Guided Testing and Automated Repair:} \system support a form of guided UI testing in which it automatically generates test cases from the abstraction pipeline and simulates learner (end-user) interactions within the final simulation UI. This feature was informed by our user study (Section~\ref{sec:userstudy}). For instance, as seen in Figure \ref{fig:widget-ux}b, in the buoyancy simulation, the system executes scripted interactions---such as adjusting the temperature slider or releasing the balloon---and prompts for Mr. Carlos's feedback. Specifically, Mr. Carlos is asked to evaluate whether the observed behavior aligns with his instructional intent, such as whether increasing temperature causes the balloon to rise faster. His response indicates whether the test has passed or failed, enabling it to refactor the code to fix any issues through human feedback. 

\subsubsection{Manual Debugging and Repair}
Beyond guided testing and automated repair, Mr. Carlos can directly inspect the simulation on his own and identify points of misalignment between intents and simulation behavior. Even with state-of-the-art models, underspecification in earlier stages can result in missing or incorrect logic. \system provided affordances for interactive debugging and refinement through a rich chat based interface along with direct annotations on the generated simulation and interaction graph (Figure~\ref{fig:screens}f). For instance, Mr. Carlos can circle around a region of interest on the simulation and inspect the relevant nodes in the interaction graph which is filtered automatically based on the annotation, and then use natural language to describe the problem. For example, he might observe that adjusting the weight slider does not affect the balloon's altitude, and describe the correction in terms of missing connections between relevant concepts.

To support such targeted correction, \system implements an underspecification resolution engine that interprets the context of his chat message and generates prefilled widgets that represent candidate fixes at the appropriate level of abstraction. Mr. Carlos can confirm, edit, or reject these suggestions, allowing him to refine the simulation without needing to modify code directly. Or, Mr. Carlos can manually select an abstraction and modify it, as seen in Figure \ref{fig:widget-ux}a. All changes are shown as drafts until he chooses to commit or discard them, enabling iterative refinement grounded in his instructional intent.

%% file: 04_system.tex
\section{System Architecture}

\begin{figure*}
  \centering
  \includegraphics[width=\linewidth]{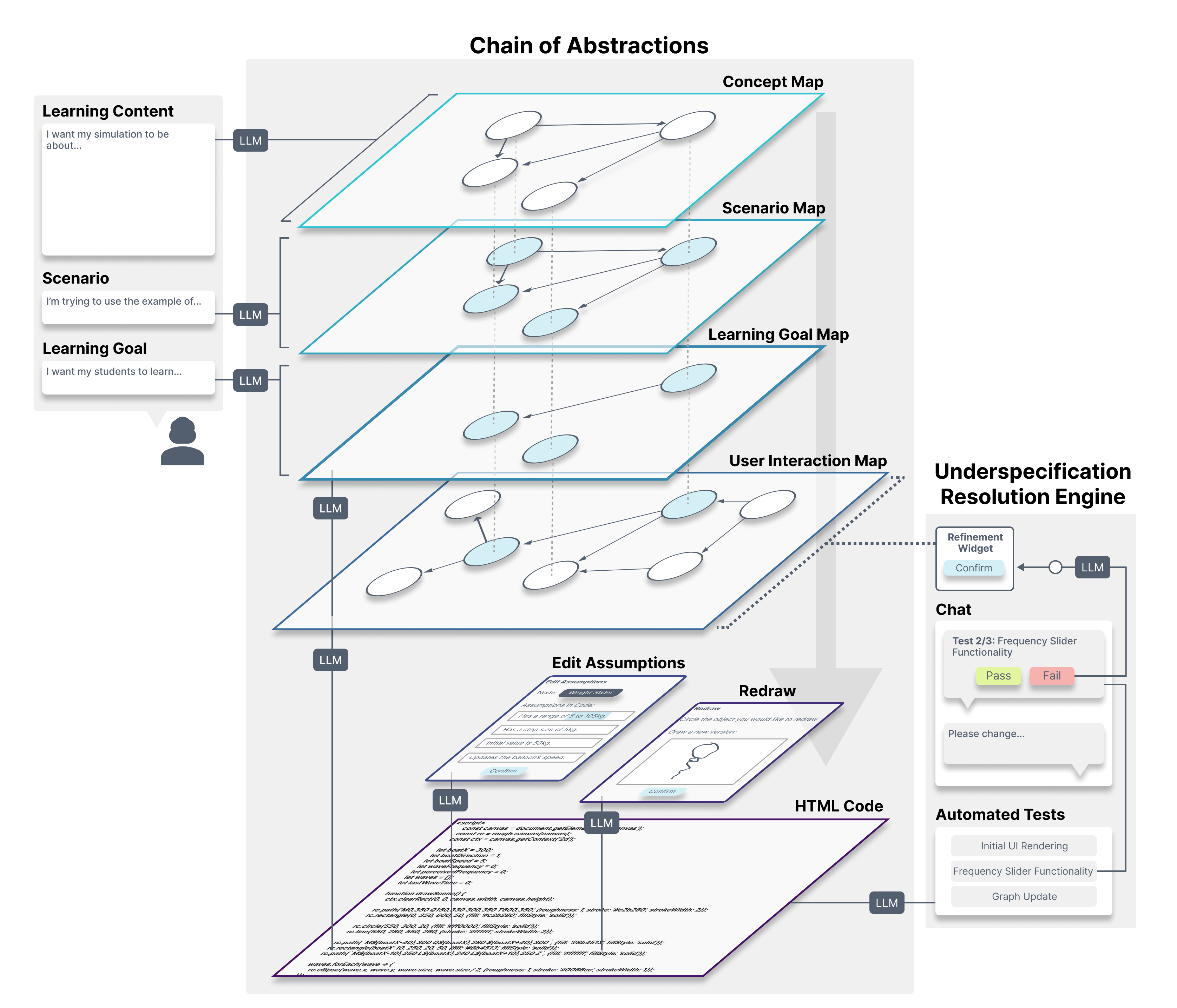}
  \caption{The system architecture of \system, consisting of a chain-of-abstractions pipeline for content generation and an underspecification resolution engine for inverse error correction. Abstractions in the chain-of-abstractions pipeline include node-link diagrams and chat-based widgets.}
  \Description{A diagram explaining the layout of \system. On the left is an icon representing the user and their chosen learning content, scenario, and learning goal. These components are connected via arrows to different abstractions within a box labeled ``Chain of Abstractions.'' This box has five abstractions with icons representing them. the first four (Concept Graph, Scenario Graph, Learning Goal Graph, and UI Graph) are all represented as horizontal layers. Two additional abstractions (Edit Assumptions and Redraw) are chat widgets. And the HTML Code is represented by a layer consisting of code. Another box labelled ``Underspecification Resolution Process'' has an input of ``Chat'' and shows automated tests generated from the code and displayed in the chat. The "Fail" button of a chat-displayed widget is connects to a refinement widget, when updates an abstraction.}
  \label{fig:architecture}
\end{figure*}

Here we describe the technical details for (1) the CoA pipeline, (2) the undespecification resolution approach, (3) automated code testing, and (4) affordances for referential conversational interactions with the LLM. 

\subsection{Chain-of-Abstractions Pipeline}
Rather than assuming educators have the necessary design experience to generate effective simulations, \system employs a chain-of-abstractions (CoA) technique that allows teachers' design decisions to be easily integrated into the previous steps of the simulation design process. Figure~\ref{fig:architecture} shows the abstractions used in this chain. \systems CoA forward abstractions are node-link diagrams representing process and UI information. \system's set of forward transformation abstractions is \[\mathcal{A} = \{\text{Concept Graph}, \text{Scenario Graph},\]\[\text{Learning Goal Graph},\text{User Interaction Graph}\}\].

\system's set of transformations $\mathcal{T}$ therefore includes five transformation, one for the generation of each abstraction in $\mathcal{A}$ and one for generation of $C$. All abstractions in $\mathcal{A}$ can be refined through a set of six direct manipulation widgets. These forward abstraction refining widgets and their implementations are outlined in Table~\ref{tab:widget-defs}.

\subsubsection{Concept Graph}

The teacher's initial learning content prompt is translated into a knowledge graph, or node-link diagram, via LLM prompting. We call this  knowledge graph abstraction the ``Concept Graph,''. In the Concept Graph, objects in the input text are become nodes, and the relationships between objects become directed links between nodes. This graph is visually displayed to the user upon generation. Knowledge graphs like this are commonly used for the representation of concepts in an educational setting, so teachers will be familiar with this form of abstraction~\cite{agrawal2022building, chen2018knowedu}.

When prompting for this graphical abstraction, \system uses natural language request (including user inputted learning content) along with a list of requirements for the form of the generated graph.

\subsubsection{Scenario Graph}

Once an initial Concept Graph is generated, the teacher then selects a scenario. \system generates the Scenario Graph by promting an LLM to update the nodes in the Concept Graph to be specific to the chosen scenario. Links are remain the same. Narrowing a simulation down to a specific scenario or example does not change the conceptual relationships presented in the simulation, but may change the objects that the simulation is engaging with. 

For example, a Concept Graph representing the states of matter may state that the nodes \framebox{Solid} and \framebox{Liquid} are connected via the link $\longrightarrow melting \longrightarrow$. If the user selects the scenario ``The Water Phase Transition'', \system would generate a scenario graph where \fcolorbox{black}{skyblue}{Ice} is connected to \fcolorbox{black}{skyblue}{Water} via $\longrightarrow melting \longrightarrow$.

\subsubsection{Learning Goal Graph}

The Scenario Graph can be complex, with many nodes and links that are not relevant to the student's hypotheses-verification process. In the end simulation, students should have the ability to prove or disprove hypotheses to achieve scenario-specific learning goals without grappling with extraneous information. To address this, \system lets the user select a scenario-specific learning objective, then translates the Scenario Graph to a Learning Goal Graph by prompting an LLM to remove any nodes and links that are unnecessary to the chosen learning goal.

\subsubsection{User Interaction Graph}

The User Interaction Graph (UI Graph) represents the full simulation, including conceptual information about the content, UI elements, and visuals. It is generated once the teacher has made all three necessary simulation design decisions (Learning Content, Scenario, and Goals).

The UI Graph generation process involves identifying key objects and relationships in the learning goals and generating an experimental procedure based on this information.

Simulations addressing \textbf{descriptive knowledge} generate their UI Graph by first identifying (1) the independent variable of the learning goal, (2) the dependent variable, and (3) the relationship between these two variables. The third characteristic is identified as the learning goal itself, but \systems process also prompts for the independent and dependent variables. Then all these are characteristics are used to generate an experimental procedure through LLM prompting. Finally, \system prompts to translate this procedure into an UI Graph with all necessary interactions to complete this procedure. For instance, if the learning goal is to understand how sunlight affects the height of plants, the procedure will involve the independent variable --- amount of sunlight, the dependent variable --- the height of the plant, and the relationship: more sunlight leads to taller plants. The specific procedure might be as follows:

\begin{promptbox}
    \begin{lstlisting}
        (1) Select the plant for observation.
        (2) Expose the plant to full sunlight for a week and measure its height daily.
        (3) Change the condition to partial sunlight for the next week and continue measuring its height daily.
        (4) Reduce sunlight to no sunlight for the final week and again measure its height daily.
        (5) Record and compare the data to determine how the different sunlight conditions affected the plant's growth.
    \end{lstlisting}
\end{promptbox}

For \textbf{explanatory knowledge}, we find four main characteristics that must be identified. First, \system prompts for the main experimental object that the learning goal is exploring. Then \system again identifies the independent and dependent variables of the learning goal, along the explanatory object, or the object that explains the relationship between the independent and dependent variables. All four of these characteristics are then used to prompt for an experimental procedure, which is then translated to the UI Graph.

Learning goals that address \textbf{procedural knowledge} use two main characteristics. Namely, \system prompts for both the main experimental object that the learning goal is investigating, along with the underlying process that is explained through the learning goal. Using both of these features, \system then generates an experimental procedure to uncover the process presented in the learning goal. And again, this is translated into a UI Graph including all necessary experimental objects and processes.

\subsubsection{Simulation Code}

The UI Graph is then directly translated into simulation code. This simulation code includes HTML, CSS, and JavaScript for all functionality and visual elements included in the UI Graph. 

\begin{table}
\begin{tabular}{ |p{3cm}|p{5cm}|  }
 \hline
 \multicolumn{2}{|c|}{\system Forward Abstraction Refinement Widgets} \\
 \hline
 Widget&Description\\
 \hline
 Add Node&Requires the user to input the name of their new node and adds that node to the current abstraction.\\
 \hline
 Add Link&Requires the user to draw a new link between two nodes and input the label of their new link. Adds a new link between the provided nodes with the provided label to the current abstraction.\\
 \hline
 Remove Node&Removes the selected node from the current abstraction.\\
 \hline
 Remove Link&Removes the selected link from the current abstraction.\\
 \hline
 Edit Node Label&Changes the label of the selected node to a new, user inputted label.\\
 \hline
 Edit Link Label&This widget changes the label of the selected link to a new, user-inputted label.\\
 \hline
\end{tabular}
\caption{The widgets used to refine forward abstractions in \system.}
\label{tab:widget-defs}
\end{table}

\subsection{Underspecification Resolution Approach}
\label{sec:ure}
In \system, inverse correction is implemented using the Underspecification Resolution Engine. This process allows the user to correct any assumptions made by the LLM revealed at the final code level by refining abstractions in $\mathcal{B}$:

\[\mathcal{B} = \{\text{UI Graph},\text{Code Assumptions Abstraction},\]\[\text{Redraw Abstraction}\}\]

The \textbf{UI Graph} maintains the same forward operation implementations as in the forward pass. In the \textbf{Code Assumptions Abstraction}, the user uses a chat widget to select a node in the UI Graph and inspect a list of assumptions that the code is making about that node. The user can then refine these assumptions through text editing. Once they've made edits, \system prompts an LLM to correct the code's implementation to reflect these updated assumptions. And the \textbf{Redraw Abstraction} requires the user to circle an object in the end simulation and create a rough sketch of what they want that object to look like visually using a chat widget. The system then prompts an LLM to update the circled visual to more closely match the user's rough sketch.

Using these abstractions, \system implements two approaches to underspecification resolution. 

\subsubsection{Guided Testing and Automated Repair}

\begin{figure*}
  \centering
  \includegraphics[width=0.7\linewidth]{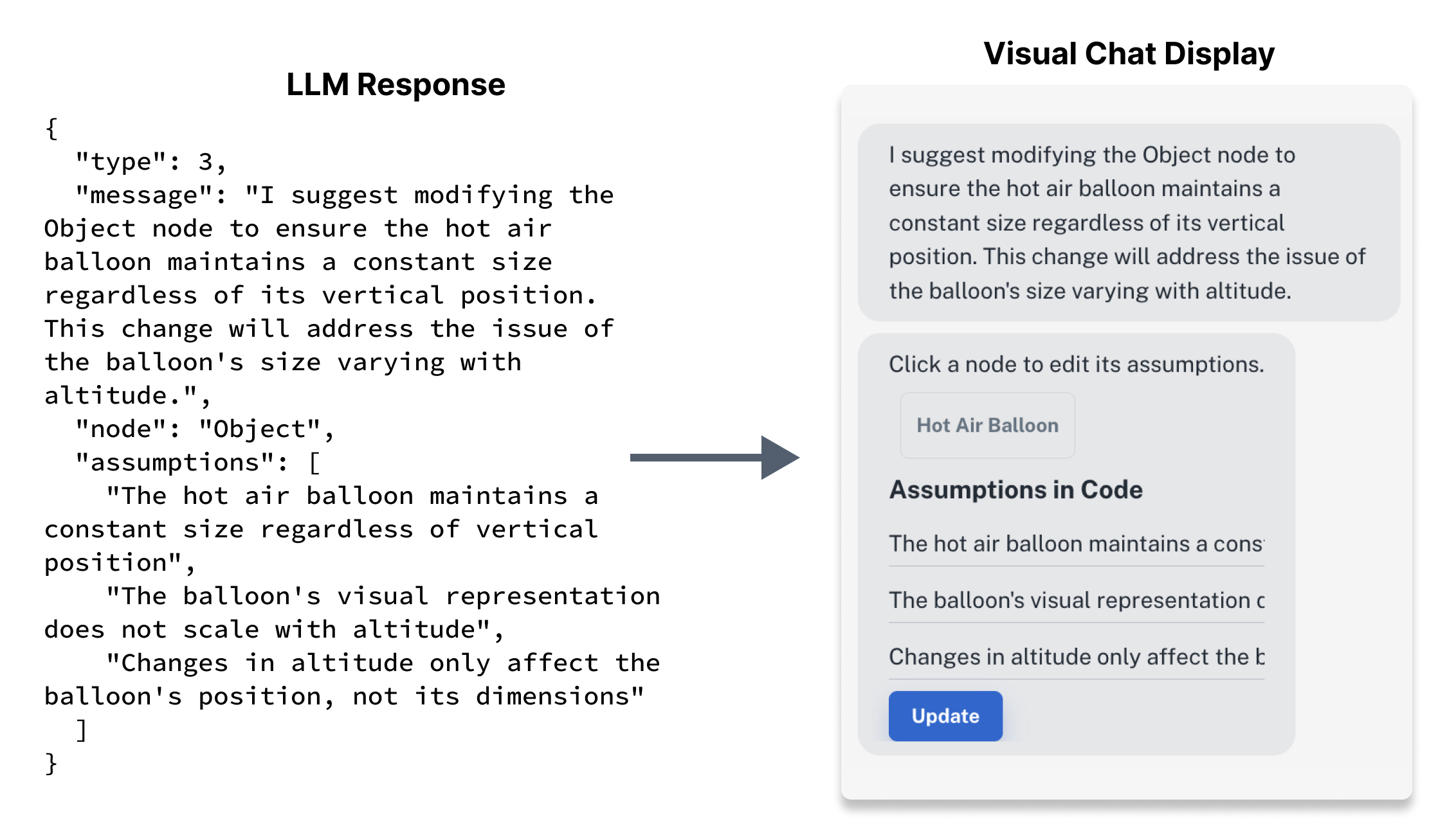}
  \caption{The JSON LLM response representing the "Edit Assumptions" abstraction, along with its visual display in the chat. "type" in the JSON object indicates the type of chat widget returned. Note that the "node" attribute corresponds to the ID of the relevant node, while the value displayed in the widget displays the label on that node.}
  \Description{A diagram where on the left there are two visuals, one representing an original HMTL code block and, below, another representing an updated HTML code block. Overlayed with the original code block is a chat message reading "Make the weight slider range from 5 to 105kg." This chat message is expanded on the right to show the same underspecification resolution engine diagram as in Figure~\ref{fig:architecture}, except with the "Edit Assumptions" widget chosen. This assumptions widget is also displayed with completed information, and is then connected to the updated HTML code block, which highlights the changed code.}
  \label{fig:prompt-to-chat}
\end{figure*}

This approach combines automated code testing and abstraction modification suggestions. As described in Section \ref{sec:automated-testing}, \system automatically tests and resolves code issues. However, tests involving user interface components are displayed to the user as chat widgets instead of automatically being resolved. Using these widgets, the user can automatically "play" UI actions and assess whether they produce expected results. When the user indicates that an unexpected result has occurred, \system invokes inverse correction. \system uses an LLM to select the previous abstraction in $\mathcal{B}$ that is most closely aligned with the assumption of interest, then automatically inspects and refines that abstraction. All the user must do is validate the updated abstraction $B_i'$. This updated abstraction is shown to the user using a chat widget..

In this approach, a pre-filled assumption modification widget is returned from the LLM as a JSON object with all necessary information. See Figure~\ref{fig:prompt-to-chat} for an example of this response for the ``Edit Assumptions'' abstraction.

\subsubsection{Manual Debugging and Repair}

The user can also manually select an abstraction from $\mathcal{B}$ to inspect, refine, and validate. For example, if the user notices that the concept graph is missing a node, they can directly select and fill out the "Add Node" widget to add it in. Or they can prompt the chat explaining the error, in which case an LLM will refine the affected abstraction for the user. The user can verify and accept this refinement.

\subsection{Automated Code Testing}
\label{sec:automated-testing}

\begin{figure*}
  \centering
  \includegraphics[width=\linewidth]{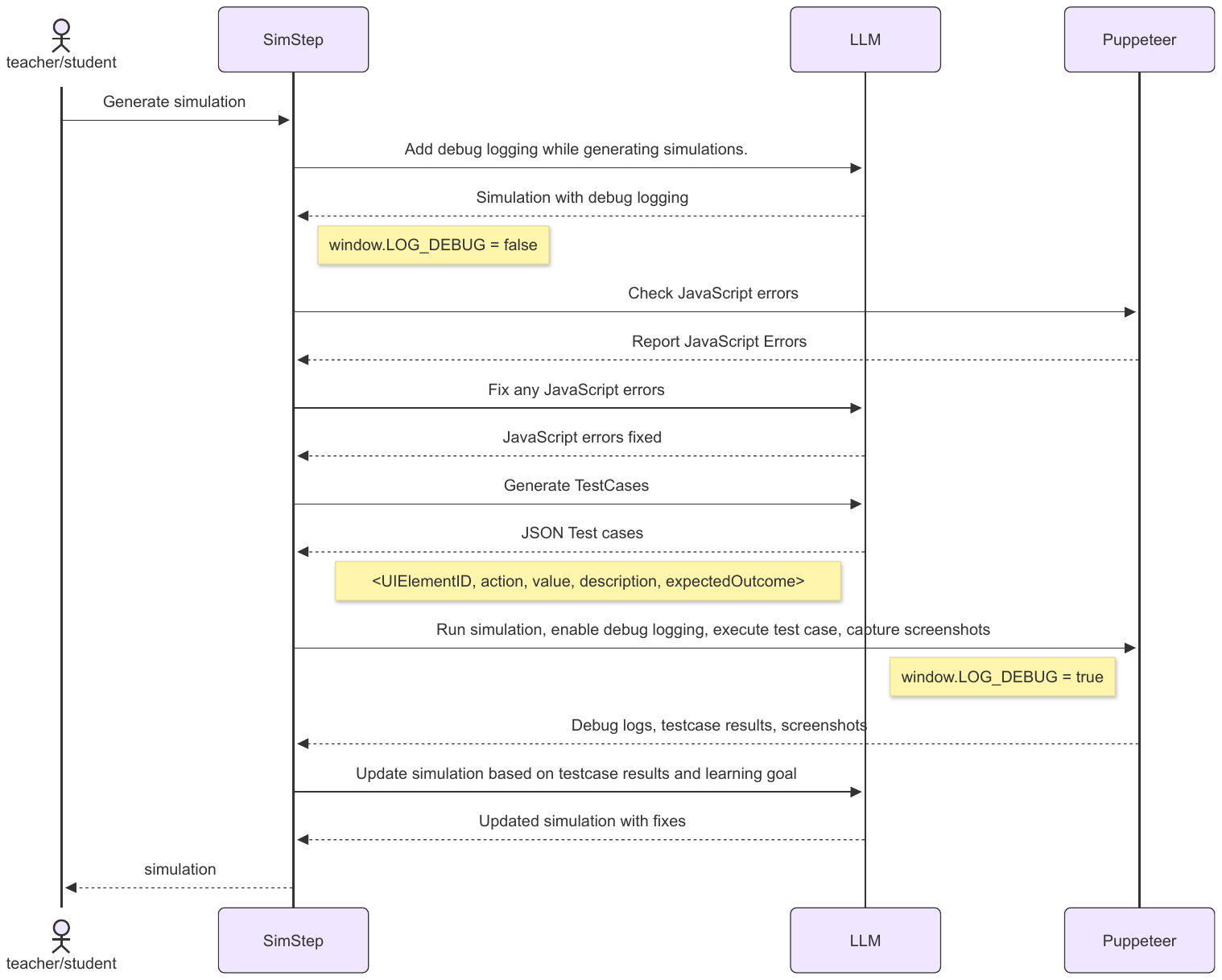}
  \caption{Automated Testing and Error Resolution Workflow for Simulation Verification}
  \Description{This diagram outlines the end-to-end process for verifying and fixing simulation functionality through automate testing. The workflow begins by Puppeteer detecting any JavaScript errors and prompting the LLM to fix them. Once the simulation is free of JavaScript errors, the LLM generates structured test cases based on the UI context and learning goals in a JSON format. Puppeteer then executes these structured test cases by capturing screenshots and detailed debug logs for every action. This info is passed to the the LLM which verifies the test results and, if any test case fails or the learning goal is not met, it automatically updates the simulation code to resolve the identified issues.}
  \label{fig:testing-sequence}
\end{figure*}

We noticed that some of the generated simulations do not work due to JavaScript, logical or User Interface issues. In order to fix this, we use a headless browser approach, combined with detailed logging. The detailed logging is achieved by capturing the UI and logic state of the simulation in the form of log messages, as well as taking screenshots after UI actions. We automate this using Puppeteer, a Node library used to control headless Chrome.  Figure~\ref{fig:testing-sequence} shows the sequence diagram for the automated testing workflow, which happens seamless to the user.

\subsubsection{JavaScript error resolution}
We first ensure that the simulation has no JavaScript errors. This is crucial since JavaScript errors can prevent the UI from working properly. Since button clicks could also result in JavaScript errors, we go through all buttons and perform a click action for every button. All JavaScript errors are then captured using Puppeteer and sent to the LLM to fix. 

\subsubsection{Test case generation}
Once the JavaScript errors are resolved, we ask the LLM to generate test cases. Each test case is defined as a JSON object with the following structure:

\begin{lstlisting}[style=compactjson]
// Identifier of the UI element
"ID": "slider-weight",
// Action to perform
"action": "set_value",
// Value to set
"value": 80,
// Description of what's being tested
"description": "Adjust weight to observe balloon response",
// Expected outcome
"expectedOutcome": "Balloon altitude decreases",
// Whether this is UI-specific
"isUIVerification": true
\end{lstlisting}

\subsubsection{Automated test case execution and verification}
\label{sec:automated-testing}
For tests that do not relate to UI components, \system automatically executes, verifies, and updates the code based on the test. In order to provide a comprehensive context to the LLM, we enable debug logging (e.g., capturing SVG coordinates and action timestamps) before executing the test cases with Puppeteer. We run each test case by executing the specified action and capturing a screenshot at the end of each test step. We also capture an initial and final screenshots of the simulation.

The LLM is then invoked with the test case execution results from Puppeteer, the simulation code and the learning goal. We ask the LLM to verify each test case and update the simulation code if the tests fail or if the learning goal is not met. If the LLM finds that a test case is not successful, or a learning goal is not satisfied, it will attempt to update the HTML with an updated version that addresses the underlying problem. We found that it does a very reliable job of ensuring logical errors related to the simulation are fixed. The detected UI errors are fixed based on the limited vision capabilities of the LLM.

Tests involving UI components are displayed in the chat for the user to verify.

\subsection{Improving Collaborative Interactions Between LLM and User}

A side effect of the CoA is that the user must keep maintain understanding of multiple, often complex, abstractions of their authored simulation. This can make communicating with the LLM during the Underspecification Resolution Process cognatively expensive. As a means of improving the user experience of \system during this process, we have also implemented features that allow users to more clearly communicate their ideas to the LLM, along with allowing users to understand the assumptions of the LLM.

\subsubsection{Chat Add-Ons}

When prompting via the chat, the user often would like to reference simulation abstractions to explain their desired changes. In order the aid the user in doing this, the user can draw annotations on top on the current simulation. Every time the user annotates, the annotation is labeled. In chat, the user can type "@" to get a list of all nodes in the UI Graph, along with all annotations on the simulation. The user can select from this list to reference a specific annotation or node. When prompting for the desired change to the code, the LLM is provided with these annotations and nodes as context.

\subsubsection{Connecting Code and Abstraction}
The translation from UI Graph to HTML Code can be cognitively intense for users. In order to help users understand the connection between the two abstractions, we've implemented a ``Subgraph Selector'' tool, in which the user can circle one or multiple sections of the end simulation and the tool will display the sub-graph of the UI Graph that this region corresponds to. This is implemented by prompting the LLM with an image of the annotated simulation along with the context of the simulation code and UI Graph.

\subsection{Implementation Details}

\subsubsection{Frontend}
The frontend of the \system application is built using \texttt{React}~\cite{react}. For many of the UI elements and visual components, we use Material UI~\cite{mui}, a UI component library. In order to visually and directly manipulate the abstractions that have a graphical structure, we use \texttt{joint.js}~\cite{joint}. While for parsing these graphs, we use a \texttt{mermaid.js}~\cite{mermaidjs} format. We also employ \texttt{tldraw}~\cite{tldraw} for annotating in the Interactivity Page. 

\subsubsection{Backend}
The backend is built using Node.js~\cite{node} and Express.js~\cite{express}. We use Anthropic's Claude~\cite{anthropic2023claude} claude-3.5-sonnet model for all large language model prompting, which we do throughout the process of generating and modifying abstractions in our system. In the design of prompts in this system, we employed several prompting techniques in a trial-and-error approach, resulting in outputs that have consistent form and quality. We also store user-generated simulation code in a Firebase~\cite{firebase} database. This allows users to access the simulations they've created by link.

\subsubsection{Prompt Engineering}

By aligning the interface with the strengths and limitations of the LLM, we created a system that simplifies the authoring process while maintaining the necessary balance between AI capabilities and user control. As Adar puts is \textit{``Your UI shouldn’t write checks your AI can't cash, and your AI shouldn’t write checks your UI can't cash\footnote{https://www.youtube.com/watch?v=11UKXaELg8M}.''}.

%% file: 05_user_evaluation.tex
\section{User Evaluation}\label{sec:userstudy}
To understand the user experience of \system's Chain-of-Abstractions (CoA) approach, we conducted a user study with $N = 11$ educators. Participants had an average of 9.18 years of teaching experience ($\mu = 9.18$, $\sigma = 6.98$) and were recruited through social media and the researchers’ professional networks. All participants had prior experience teaching STEM subjects, spanning grade levels from middle school to undergraduate college. Each study session lasted approximately 90 minutes and was conducted one-on-one over Zoom, during which participants interacted with a deployed version of \system. Participants received a \$40 honorarium for their time. The study protocol was approved by the institution's Institutional Review Board (IRB).

\subsection{Method}
In each session, one of the researchers gave a participant a 15-minute demo of \system, going through the process of generating a simulation about \textit{States of Matter and Phase Change}. The participant could ask any question they had about the tool. Then, the researcher allowed the participant to use \system to generate their own simulation about either a subject matter of the participant's choice or one of two pre-prepared learning content (Gravitational Force and Buoyancy). When time permitted, a second simulation was generated using a second learning content. After the participant generated and corrected at least one simulation, they were asked to reflect on their experience using \system through an unstructured interview. Finally, participants filled out a questionnaire on the usability of the system, the task load, and the Cognitive Dimensions of Notations~\cite{green1989cognitive}. The usability of the system was evaluated using the Post-Study System Usability Questionnaire (PSSUQ)~\cite{lewis1992psychometric}, and the task load was assessed using the NASA
Task Load Index (NASA-TLX)~\cite{hart1988development}.

\subsection{Results} 
\hfill

\begin{figure}
  \centering
  \includegraphics[width=\linewidth]{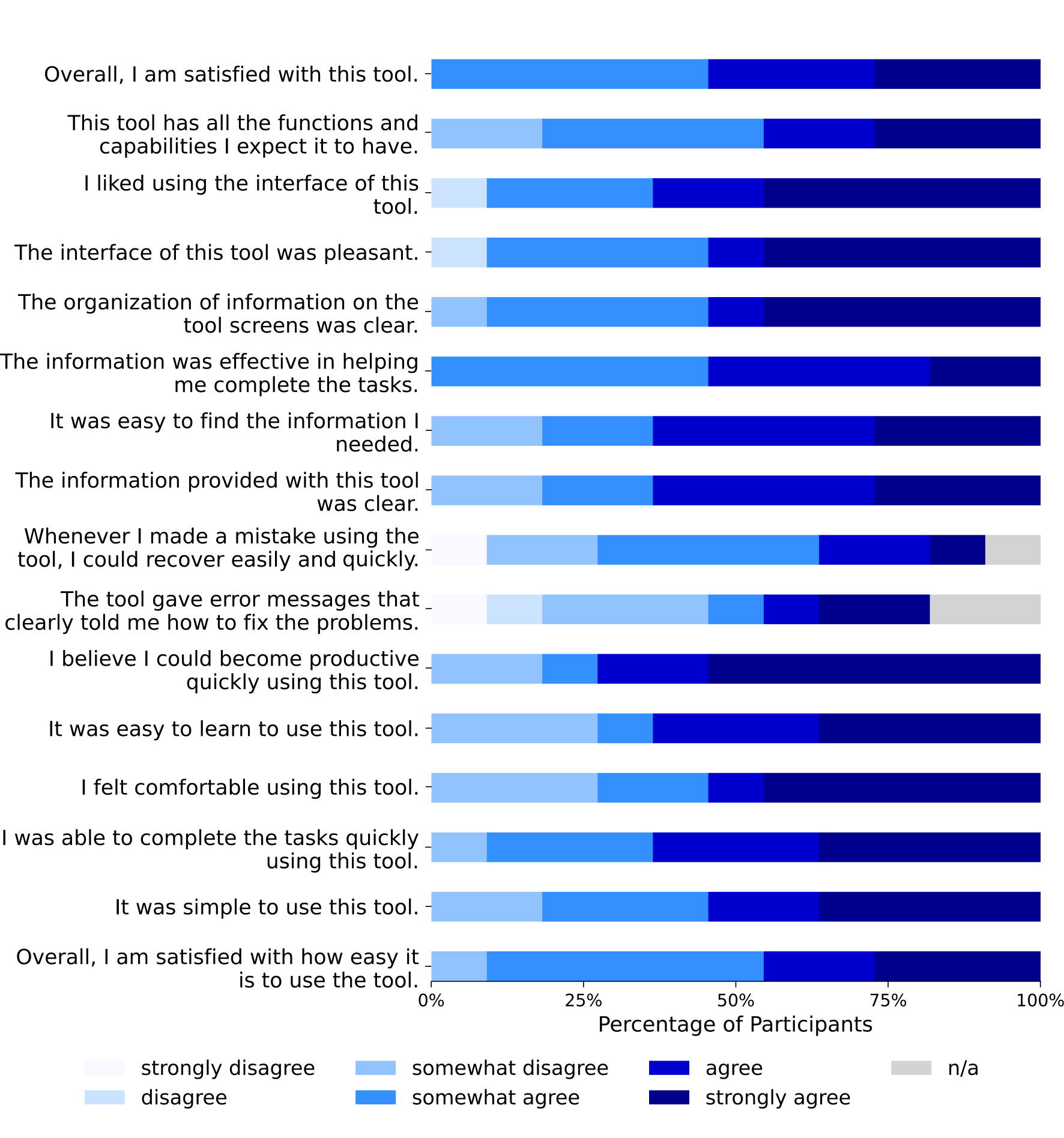}
  \caption{Teacher Participant Responses to Post-Study System Usability Questionnaire (PSSUQ)}
  \Description{A horizontal stacked bar chart with a bar for each question in the PSSUQ. Below, there is an legend that explains the meaning of each bar color (each corresponds with a type of participant response, such as "Agree" or "Strongly Disagree").}
  \label{fig:system-usability}
\end{figure}

\begin{figure*}
  \centering
  \includegraphics[width=\linewidth]{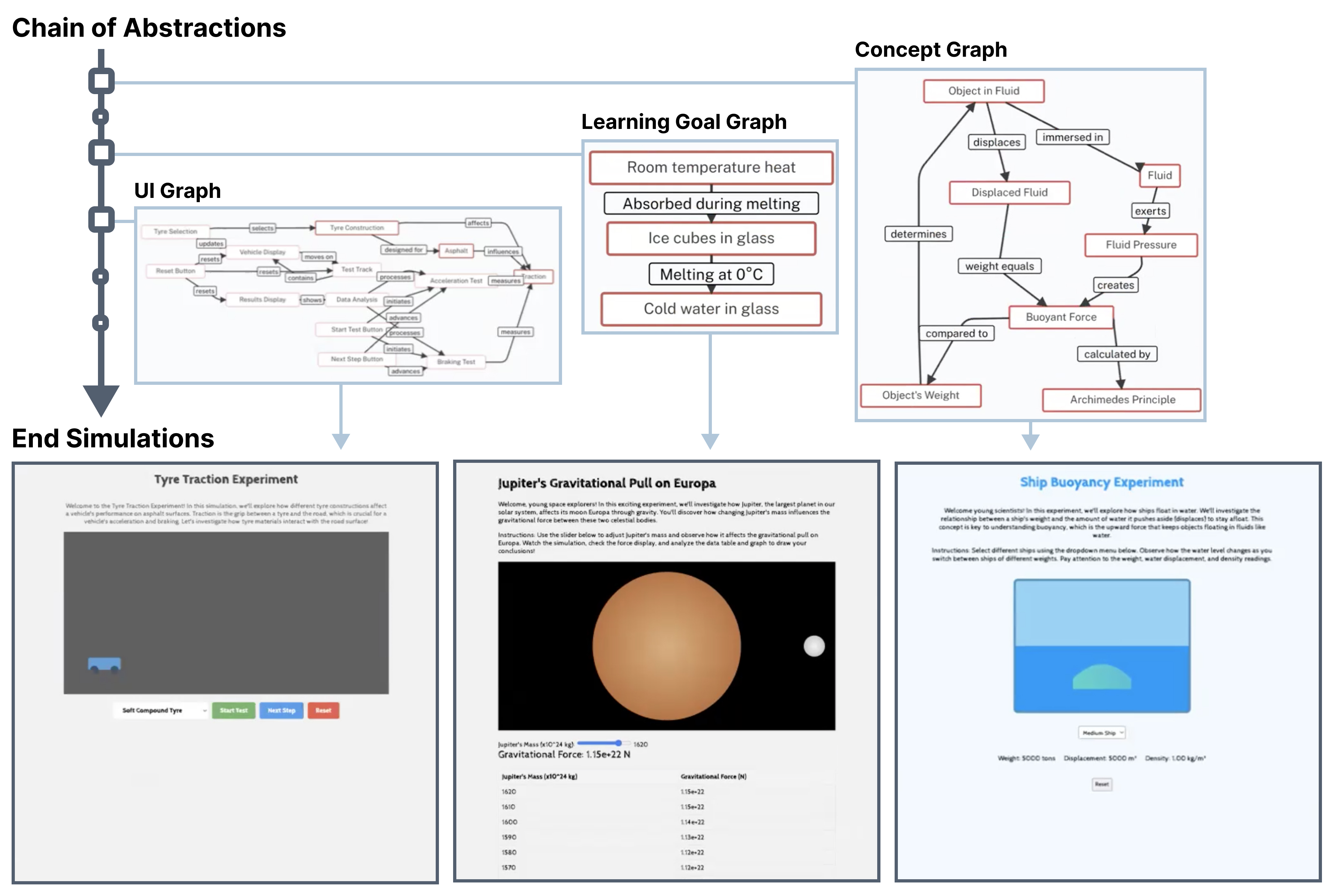}
  \caption{A selection of simulations and intermediate representations generated by teachers during our expert evaluation study. This set of simulations includes an exploration of friction on different tire materials, mass versus gravitational force, and buoyancy on different sized ships.}
  \Description{Three screenshots of simulations arranged horizontally. Left to right the simulations are: (1) a simulation depicting a boat in water titled "Ship Buoyancy Experiment," (2) a simulation depicting two planets titled "Jupiter's Gravitational Pull on Europa," and (3) a simulation depicting a blue car titled "Tyre Traction Experiment." Above each screenshot is a previous abstraction in the CoA: (1) UI Graph, (2) Learning Goal Graph, and (3) Concept Graph. Each is connected to their corresponding simulation with an arrow.}
  \label{fig:teacher-sims}
\end{figure*}

\subsubsection{System Usability} Overall, teachers found \system intuitive to use and reported that they would feel comfortable using this system in a classroom setting. In the PSSUQ questionnaire, we anchored 1 as ``Strongly Disagree'' and 6 as ``Strongly Agree.'' Our overall usability score was 4.66 ($\sigma$ = 0.36), and in qualitative feedback multiple participants commented that the system felt ``natural to use.'' The System Usefulness (SYSUSE) sub-scale was 4.8 ($\sigma$ = 0.15), indicating that the teachers viewed this tool as useful to their teaching process. Teachers also remarked that they could see themselves using this tool in a variety of ways, including to create introductory material for students, exploratory expansions of their class subjects, and replacement materials for in-class activities such as experiments. Further, the Interface Quality (INTERQUAL) sub-scale had a score of 4.78 ($\sigma$ = 0.15), indicating the teachers felt that they had the ability to navigate and alter the CoA and the underspecification resolution process with ease. However the Information Quality (INFOQUAL) sub-scale had a score of 4.44 ($\sigma$ = 0.48), indicating that we can improve the quality of information displayed to the user. As seen in Figure~\ref{fig:system-usability}, several users found that error messages were not present or useful during their process. There is room for integration of more intuitive error messages when the user encounters an unexpected error in the system. 

\begin{figure}
  \centering
  \includegraphics[width=\linewidth]{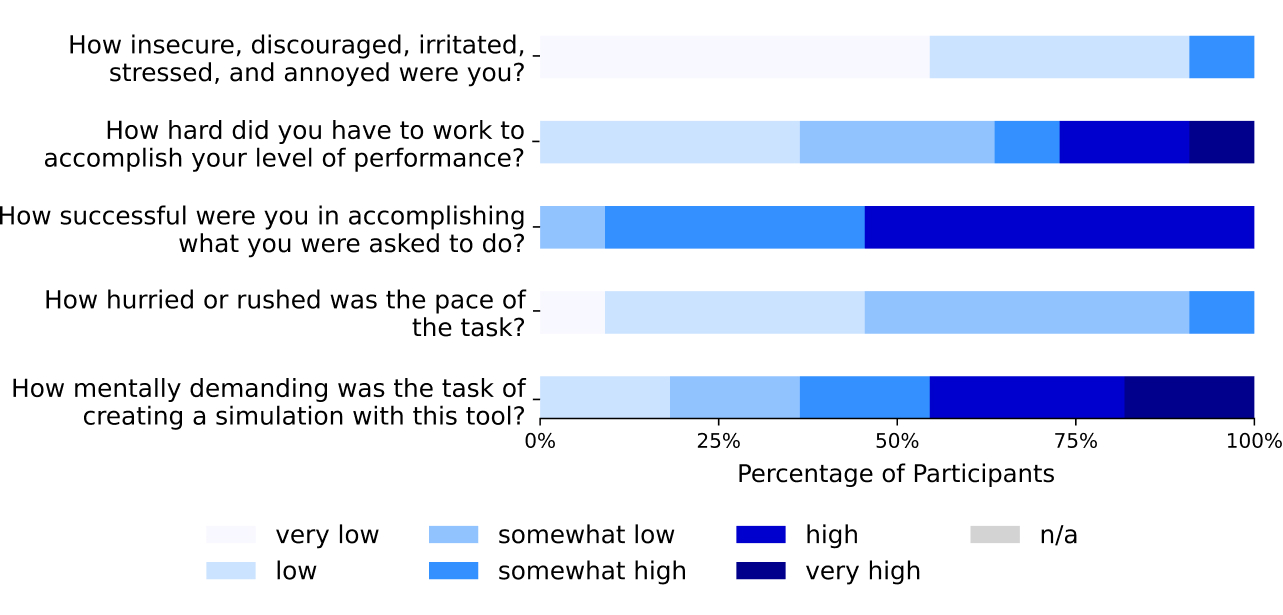}
  \caption{Teacher Participant Responses to NASA's Task Load Index (NASA-TLX) Questions}
  \Description{A horizontal stacked bar chart with a bar for each question in the NASA-TLX. Below, there is an legend that explains the meaning of each bar color (each corresponds with a type of participant response, such as "low" or "very high").}
  \label{fig:task-load}
\end{figure}

\subsubsection{Task Load}
Although \system does not eliminate the task load of simulation generation, our system provides an interface that imparts a load onto users that is not high. We evaluated task load using NASA-TLX with a 1-6 scale. We anchored 1 with ``Very Low'' and 6 with ``Very High.'' We did not directly assess the physical demand of our tool considering digital nature of our tool. The unweighted TLX score for \system was 2.64, indicating that although teachers felt that the task of generating a simulation required work, this load was not high. In Figure~\ref{fig:task-load}, we can see that teachers felt that there was considerable mental demand in the simulation generation process. But regardless on this mental demand, most participants felt that they were ultimately successful in the process. See Figure~\ref{fig:teacher-sims} for some of the simulations that teachers generated during this study. In qualitative feedback, participants remarked that there was low mental demand in the initial steps of the simulation generation process. Mental demand increased at the step of correcting behavior and errors at the simulation code abstraction level (i.e., the Interactivity Page). Participants most often engaged in the underspecification resolution process by prompting the chat. In association with the task load of this activity, one participant remarked ``I wanted to convey my thoughts specifically enough that they would be understood, so I struggled at first with how to describe things.'' In response to participants' feelings of being lost in the under-specification resolution process, the researchers augmented the process with automated code testing and guided testing functionality, as discussed in section \ref{sec:automated-testing}.

\begin{figure}
  \centering
  \includegraphics[width=\linewidth]{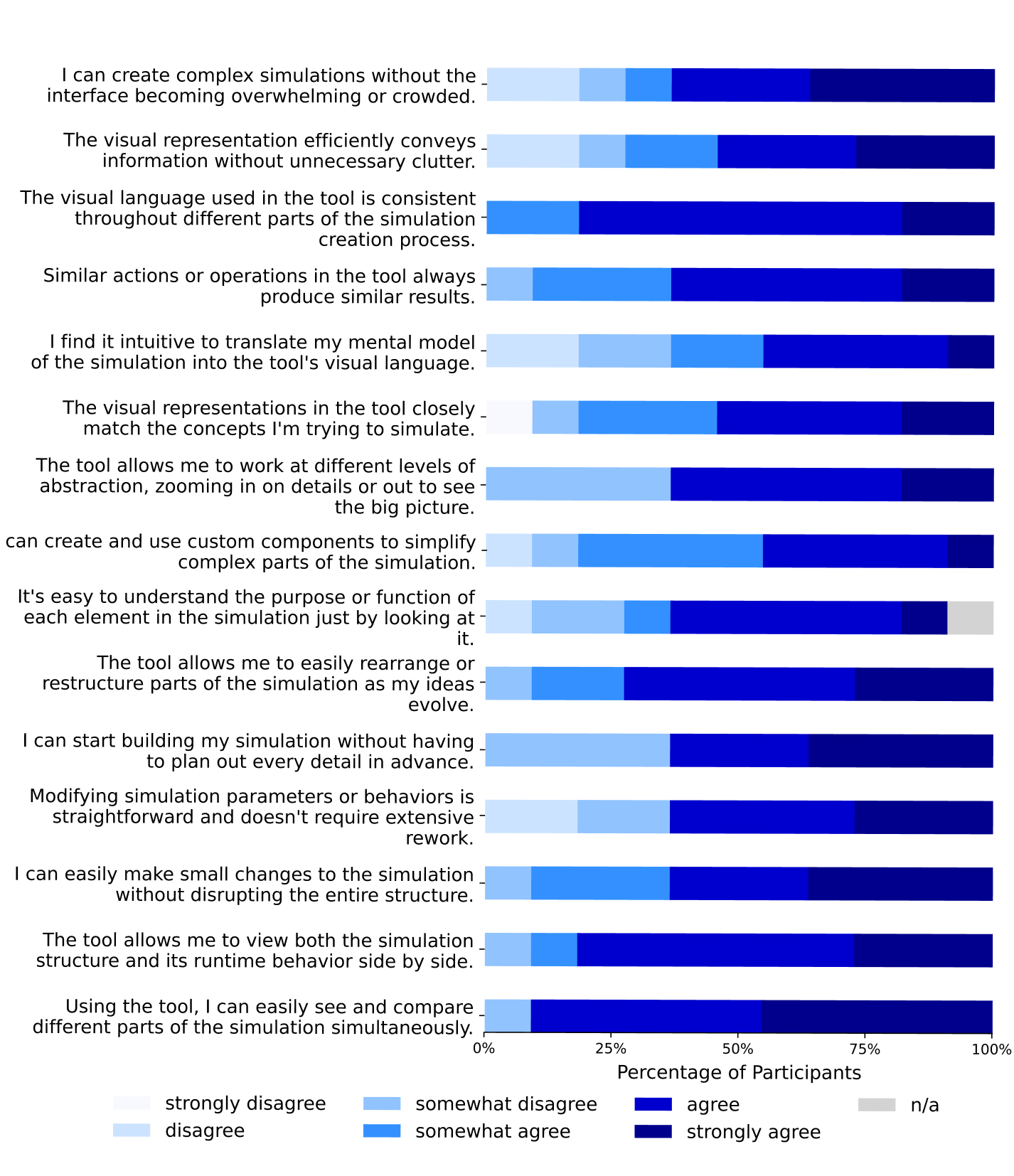}
  \caption{Teacher Participant Responses to a Questionnaire on the Cognitive Dimensions of Notations}
  \Description{A horizontal stacked bar chart with a bar for each question in a questionnaire on the Cognitive Dimensions of Notations. Below, there is an legend that explains the meaning of each bar color (each corresponds with a type of participant response, such as "Agree" or "Strongly Disagree").}
  \label{fig:cog-dim}
\end{figure}

\subsubsection{Cognitive Dimensions of Notations}
Teachers also found that the the notations used in the \systems CoA were intuitive and mirrored their own internal process when planning to teach about a subject. More specifically, we assessed these notations (abstractions) using the Cognitive Dimensions of Notations~\cite{green1989cognitive}. These dimensions allow us to evaluate whether or not these abstractions are useful to teachers for the task of simulation generation, and are often used to evaluate representational tools~\cite{bostock2009protovis, lee2023deimos}. We found that \system meets all the most relevant dimensions. In this study, we evaluated users' experiences of the system specifically for the dimensions of Visibility, Viscosity, Premature Commitment, Role-Expressiveness, Abstraction, Closeness of Mapping, Consistency, and Diffuseness using Likert scale questions with a range of 1-6. We anchored 1 with ``Strongly Disagree'' and 6 with ``Strongly Agree.'' Figure~\ref{fig:cog-dim} shows teachers' response to these questions. The average score across all included questions was 4.61 ($\mu$ = 4.61, $\sigma$ = 0.34). Questions related with visibility had a notable average score of 5.14 ($\mu$ = 5.14, $\sigma$ = 0.27), which indicated success in our goal of generating a tool that easily displays different abstractions of the same end simulation. In qualitative feedback, teachers did however indicate that they focused less on the UI Graph on the Interactivity Page, which is expected considering this page's focus on the behavior of the simulation itself.

\subsubsection{Qualitative Feedback}
In a discussion period after using the tool, participants remarked that they enjoyed the graphical abstractions of the inputted learning content. They said they found this form of abstraction to be intuitive and useful for even their own knowledge formation process. One participant remarked that \textit{``The graphical abstractions help to show the underlying concepts that are running the simulation. It is also nice to be able to adjust those which would adjust the simulation's behavior.''} Some teachers even commented that they would consider giving these maps, or the tool in general, to students to help them learning the concepts at hand. Likewise, teachers found value in the Scenario Page specifically. They appreciated that they could choose a topic specific to their students. One participant also commented that they often struggle to express learning content through different examples when students do not understand their original example. This participant said they appreciated that the tool provides example scenarios to choose, which may present the content in ways that they may not have thought of.

Despite these successes, participants did have a few suggestions regarding the visual display of the graph abstractions. The current interface does not have a zoom feature, and the generated maps can become quite complex. One participant commented \textit{``Just for me personally, I have eye disabilities, I was not able to easily zoom in on the end screen with the simulation.''} In order to address this, we plan on adding a zoom and pan feature to this display.

%% file: 06_techincal_evaluation.tex
\section{Technical Evaluation}

The effectiveness of a CoA approach in promoting distributed cognition is dependent on the \textbf{fidelity} of the included abstractions. We evaluate the fidelity of forward transformation abstractions in \system with 13 curated simulation specifications. The lead researcher collects a range of STEM learning content, culturally relevant scenarios, and learning goals using online educational resources. A learning design student then scores the generated abstractions from 1 to 10 using the question "How closely does this abstraction adhere to the previous abstraction and user inputs?", where 1 is "very far" and 10 is "very close." This student is recruited through professional connections.

\subsection{Results}

\begin{figure*}
  \centering
  \includegraphics[width=\linewidth]{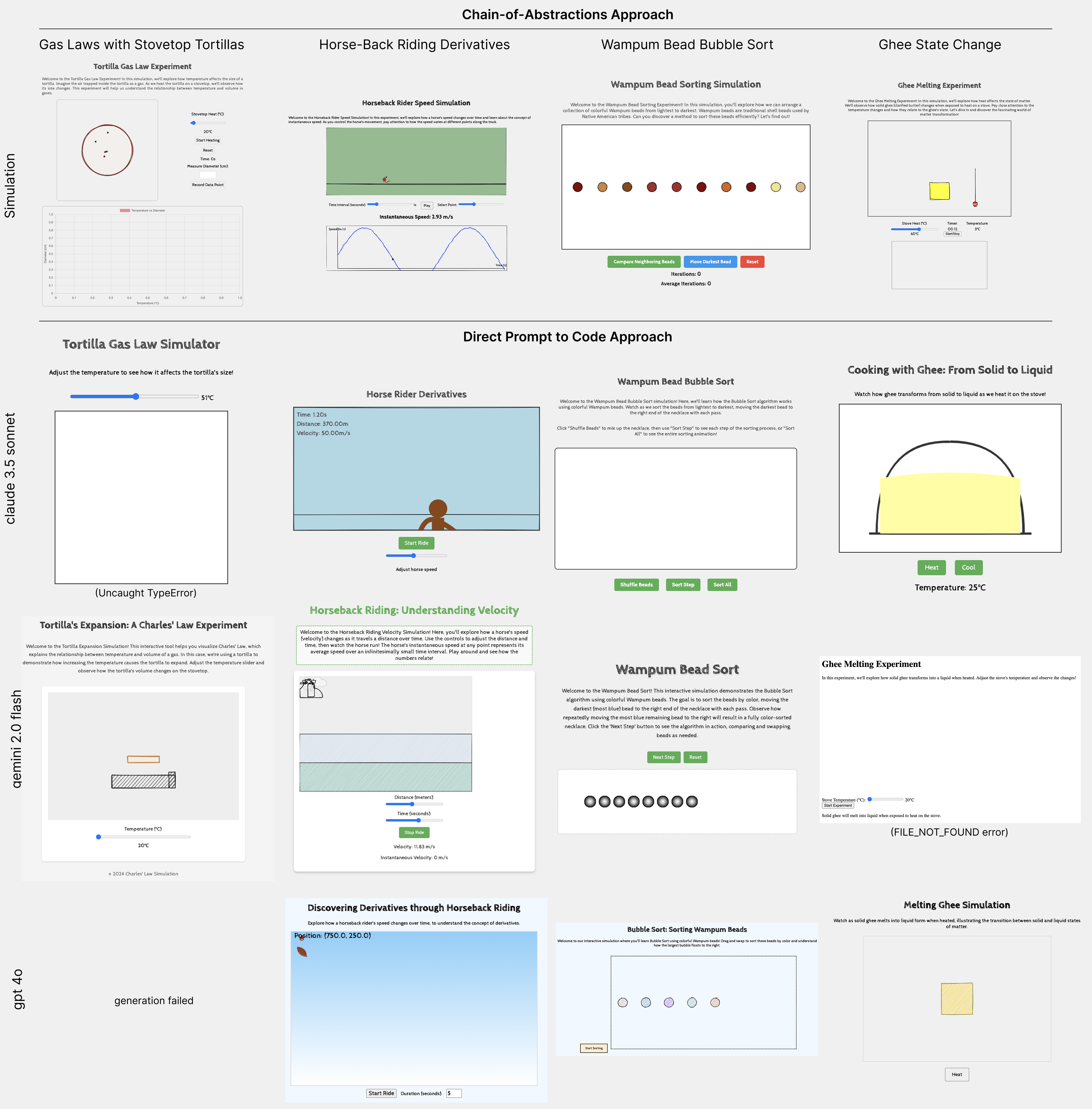}
  \caption{A selection of simulations generated through the CoA approach in comparison to simulations generated using a direct prompt-to-code approach (where the prompt includes the learning content, scenario, and learning goal)}
  \Description{A grid of node-link diagrams with four rows ("Simulation", "claude 3.5 sonnet", "gemini 2.0 flash", and "gpt 4o") and four columns("Gas Laws with Stovetop Tortillas", "Horse-Back Riding Derivatives", "Wampum Bead Bubble Sort", and "Ghee State Change"). Each cell in the grid displays a generated simulation.}
  \label{fig:eval}
\end{figure*}

In Table \ref{tab:eval}, we compare the fidelity scores of all forward transformation abstractions in \system. In general, abstractions are shown to have high fidelity, with an overall mean fidelity score of 7.6 ($\mu = 7.6$, $\sigma = 2.01$). However, \system has room to improve on both the Scenario Graph and the Learning Goal Graph abstractions, which have mean fidelity scores of 6.65 and 7.08, respectively. Future work involves performing a more in-depth study to understand the pitfalls of these abstractions and learn how to improve their fidelity by working with a range of domain experts. In Figure \ref{fig:eval}, we also compare a selection of the simulations generated in this evaluation to simulations generated using a direct prompt-to-code approach (where the prompt includes the learning content, scenario, and learning goal) with various LLM models.

\begin{table}
\begin{tabular}{ |p{3cm}|>{\centering\arraybackslash}m{2cm}|>{\centering\arraybackslash}m{2cm}|}
 \hline
 \multicolumn{3}{|c|}{\system Forward Abstraction Fidelity} \\
 \hline
 \multirow{2}{*}{\textbf{Abstraction}} & \multicolumn{2}{|c|}{\textbf{Fidelity Scores}(1-10)} \\\cline{2-3}
    & $\mu$ & $\sigma$ \\\hline
 Concept Graph&8.50&1.13\\
 \hline
 Scenario Graph&6.65&2.12\\
 \hline
 Learning Goal Graph&7.08&2.46\\
 \hline
 UI Graph&8.23&1.38\\
 \hline
\end{tabular}
\caption{All forward abstractions are scored for fidelity over 13 curated simulation specifications, and mean ($\mu$) and standard deviation ($\sigma$) fidelity scores are reported. The scorer notes that they could have potentially been biased towards assigning harsher scores to abstractions on topics that they are more proficient in.}
\label{tab:eval}
\end{table}

%% file: 07_discussion.tex
\section{Discussion}

\subsection{Desiderata for Abstractions in CoA}
In software engineering, abstractions are essential for managing complexity, modularizing functionality, and enabling reasoning at multiple levels of the system~\cite{seffah2005human}. Our CoA framework applies this perspective to programming-by-prompting by introducing intermediate representations that formalize partial intent and guide the step-by-step transformation from natural language to code. Ideally, the abstractions selected in CoA should \textbf{reflect the representational structures of the domain and the decomposition of the task}. In \system, for example, the Concept Graph captures the core domain specific knowledge and relationships, the Scenario Graph expresses contextualized learning situations, the Learning Goal Graph defines desired educational outcomes, and the UI Interaction Graph specifies how learners will interact with the system. Each of these representations corresponds to a meaningful task boundary in the teacher's workflow and provides a different lens for inspecting and refining intent. 

Across these contexts, designing effective abstractions requires attention to several key properties. In educational simulation authoring, abstractions should exhibit \textbf{visibility} by making key relationships and behaviors perceptible. For instance, a Concept Graph should clearly show that \textit{temperature affects air density}, helping a teacher reason about the cause-and-effect dynamics of buoyancy. Abstractions should also maintain \textbf{interpretive continuity} across levels; for example, a concept like ``heat'' introduced in the Concept Graph should persist through the Scenario Graph and be reflected in UI elements such as a ``heating switch'' or ``temperature slider.'' They must be \textbf{actionable}, allowing users to modify content directly such as enabling a teacher to revise the range of a slider or change the linked behavior of a button without touching the generated code. Abstractions should also support \textbf{propagation}, so that edits to a node in the Learning Goal Graph (e.g., changing ``students should understand buoyant force'' to ``students should compare hot vs. cold air'') trigger meaningful updates in downstream abstractions and ultimately in the simulation logic.

Further, the sequence of abstractions should follow a logic of \textbf{progressive formalization}. Early abstractions like Concept and Scenario Graphs align with the teacher’s domain expertise and planning practices, allowing them to express ideas in familiar pedagogical terms. Later abstractions like the Interaction Graph introduce more structure such as conditional logic or state transitions—providing a bridge between educational intent and executable behavior. At each level, users should be able to validate whether the representation reflects their goals (e.g., ``does this scenario match my intended classroom example?''), detect where the system has introduced incorrect or missing assumptions, and revise the abstraction accordingly before continuing to code generation.

\subsection{Broader Utility}
While \system exemplifies the CoA framework by helping educators author interactive simulations, the framework (Section~\ref{sec:framework}) itself has the potential to support a broad range of scenarios. The central construct of CoA is its structuring of the code generation process into a series of task-aligned abstractions. This approach connects to longstanding goals in end-user programming~\cite{Ko2011TheSOA} which aims to empower non-programmers to create, adapt, and control computational artifacts. Compared to current approaches which fall along the spectrum of tradeoffs between expressive power and ease of use, the CoA framework offers a middle path as evidenced by our user study: by representing programs as structured editable abstractions that reflect end-user's task logic, it preserves expressive power while maintaining accessibility and semantic clarity. Furthermore, our framework generalizes across domains. For instance, a data scientist might move from analysis goals to transformation logic to visual outputs~\cite{wongsuphasawat2015voyager}; a game designer might articulate core mechanics, progression rules, and interaction feedback~\cite{adams2014fundamentals}. In each case, abstractions are chosen to reflect natural task decompositions and representational practices in the domain, ensuring that the pipeline aligns with how users already think.

\subsection{Limitations and Future Work}
While the CoA framework and our implementation of \system offers a principled structure for programming-by-prompting, several limitations point towards future extensions to the framework. The current implementation, though effective for educational simulation authoring, may limit flexibility in domains where task workflows are less standardized or abstractions are harder to formalize. Designing methods for customizing or synthesizing domain-appropriate abstractions remains an open challenge. Second, while CoA reduces the need for syntactic programming, it introduces new forms of cognitive load: users must interpret and manipulate layered representations. As evidenced in the user study, the quality of outputs depends not only on the model’s capabilities, but also on the user’s ability to structure intent within the abstraction pipeline. This calls for adaptive interfaces that scaffold abstraction complexity based on user expertise. Third, current LLMs are not inherently abstraction-aware and often fail to maintain consistency across stages, suggesting opportunities for model refinement and fine-tuning. Finally, although our inverse process supports underspecification resolution, errors stemming from hallucination, misalignment, or representational gaps remain difficult to detect and correct. Ultimately, these directions point toward a broader call for understanding human-aligned representations that support not just code generation, but meaningful engagement, control, and adaptation across diverse user groups and domains.

%% file: 08_relatedwork.tex
\section{Related Work}

\subsection{Prompt Based Programming Systems}
Programming-by-prompting enables users to generate code using natural language, providing promising opportunities for end-user programming~\cite{Jiang2022DiscoveringTSA}. These systems change the role of the programmer from authoring code to verifying and debugging it~\cite{sarkar2022like}, and have shown promise in helping users solve a range of programming tasks~\cite{10.1145/3545945.3569823}. However, users frequently struggle with prompt ambiguity, underspecified behavior, and limited control over generated outputs~\cite{10.1145/3545945.3569823}. To address these challenges, recent tools embed prompting within structured workflows that offer modular prompting, multi-step interfaces, or visual scaffolds~\cite{chopra2024exploring, bouzenia2024repairagent, jiang2022discovering}. For example, Devy~\cite{bradley2018context} infers user intent from natural language to apply codebase modifications, while ProgramAlly~\cite{Herskovitz2024ProgramAllyCC} and Spellburst~\cite{angert2023spellburst} blend LLMs with domain-specific interactive UIs. These systems highlight a growing consensus that effective programming-by-prompting workflows require external representations and scaffolds to support user steerability.

Our work builds on this research trajectory by introducing the CoA framework that decomposes prompt to code generation into a sequence of structured, interpretable representations. Unlike prior systems that operate over one or two abstraction levels that are primarily program-driven, CoA defines a multi-stage pipeline explicitly designed to mirror users' task workflows and enable semantic control at each step. 

\subsection{Error Correction via Human-AI Interaction}
Even with well-formed prompts, LLM-generated code is susceptible to errors stemming from hallucinations, incomplete specifications, or semantic mismatches~\cite{austin2021program, xu2024hallucination, song2023empirical}. Research has sought to classify these errors and develop strategies for correction, including test-based evaluation~\cite{hu2024deploying}, execution-trace validation~\cite{lehtinen2024let}, and iterative refinement through self-critique~\cite{dou2024s}. Systems like Rectifier~\cite{yin2024rectifier} automate code validation using curated test suites, while Fan et al.~\cite{fan2023automated} demonstrate that program repair techniques can fix common LLM errors. Automated testing is useful, but does not reveal all hidden bugs. These testing techniques rely heavily on runtime feedback, which is not always helpful for LLM debugging~\cite{tian2024debugbench}. 

Further, these methods often treat the user as a passive recipient of model output. In contrast, hybrid systems incorporate human-in-the-loop workflows by externalizing the model's assumptions, allowing users to inspect, correct, and guide synthesis~\cite{white2024chatgpt, Liu2023WhatIW}.  Other systems such as Whyline~\cite{ko2004designing} and Hypothesizer~\cite{alaboudi2023hypothesizer} reimagine debugging as a process of explanation and hypothesis-testing, rather than syntactic error fixing. These tools allow users to query program behavior through natural language or runtime traces, reinforcing the idea that debugging can be a meaning-making activity. In this work, we combine automated testing techniques with human-in-the-loop prompting techniques, such as directly revealing assumptions~\cite{white2024chatgpt}. Further, we extend this approach through a structured \textit{inverse correction process}, which explicitly surfaces underspecifications and assumptions in code at each abstraction level in the CoA pipeline.

\subsection{Abstractions for Programming}
End-user programming systems have long used representational abstractions to help non-programmers express complex behavior~\cite{myers2006invited}. Programming abstractions come in all shapes and sizes, but abstractions including concept graphs aimed at representing high-level relationships ~\cite{Weyssow2022BetterMT},  block-based programming structures ~\cite{Zhu2020HierarchicalPF}, and scene graphs representing semantic information ~\cite{Armeni20193DSG, Ritschel2022CanGD} tend to be hierarchical and domain-specific. Early tools like KidSim~\cite{smith1994kidsim} and more recent platforms such as Spellburst~\cite{angert2023spellburst} employ block-based or visual metaphors to enable rule creation without traditional syntax.

Beyond syntactic simplification, abstractions play a deeper cognitive role: they help users approach complex tasks by engaging with structured external representations. Scene graphs~\cite{Armeni20193DSG, Ritschel2022CanGD}, concept maps~\cite{Weyssow2022BetterMT}, and scenario models~\cite{brady2022block} are examples of domain-specific abstractions that encode relationships, behaviors, and user goals in an interpretable form. These representations are not just visualization aids—they serve as manipulable scaffolds for cognitive work, compliant with the theory of distributed cognition~\cite{hutchins1995cognition}, which emphasizes how reasoning is offloaded onto and supported by external artifacts. Recent research also expands on how users engage with these abstractions in the context of authoring and debugging interactive systems. Tools like CodeToon~\cite{suh2022codetoon}, RealitySketch~\cite{suzuki2020realitysketch}, and Kitty~\cite{kazi2014kitty} demonstrate how structured workflows combined with sketching, annotation, or direct manipulation can support creative tasks. Gao et al.~\cite{gao2024efficient} explore how abstractions can similarly be applied to improve LLM accuracy and tool use by proposing a novel Chain-of-Abstraction reasoning approach.

Building on these threads, our work introduces the Chain-of-Abstractions (CoA) framework as a generalization and formalization of abstraction-based content authoring. Further, CoA emphasizes domain alignment and task decomposition, drawing from interactive content authoring systems that integrate goals, scenarios, and interaction patterns into the programming workflow~\cite{vrettakis2020story, saquib2021constructing, subramonyam2020texsketch, rosenberg2024drawtalking, cheema2012physicsbook}. As such, \system leverages CoA not only to scaffold code generation but also to support ambiguity resolution, increase control, and make the authoring process accessible to non-programmers across diverse domains.

%% file: 09_conclusion.tex
\section{Conclusion}
This work introduces the Chain-of-Abstractions (CoA) framework as a principled approach to programming-by-prompting, one that treats code generation not as a single-shot translation, but as a structured process of task-level semantic articulation. By decomposing the synthesis process into cognitively meaningful, domain-aligned representations, CoA enables users to externalize, inspect, and iteratively refine their intent. We instantiate this approach in \system, a tool that supports educators in authoring interactive simulations through a scaffolded, human-in-the-loop workflow. \systems inverse correction process addresses underspecification by surfacing assumptions and guiding revision at abstraction checkpoints, recovering key affordances of traditional programming such as traceability, testability, and control. CoA provides a foundation for more controllable, expressive, and domain-sensitive code generation.

%% file: 10_appendix.tex
\appendix

\section{Architecture Prompting}

Our chain-of-abstractions architecture utilizes LLM prompting for the generation and modification of abstractions. In this section, we document the prompts used in \system.

\subsection{Forward Abstraction Generation}

All graphical abstractions are represented as mermaid.js graphs. In prompts that include the graphical structure outputs, we specify this format:
\begin{promptbox}
\begin{lstlisting}
Use mermaidjs. The format of the mermaidjs graph should have each node with a label surrounded by square brackets, and each link with a label surrounded by vertical bars. Each line defining a node or link should begin with 4 spaces. Don't add any addition styling nor any blank lines. More specific instructions: The diagram must start with graph LR (or graph TD). Each node should be defined as NodeID[Node Label] with a single space before it. Each edge should be defined as SourceID -->|Edge Label| TargetID without any extra spaces between the arrow and the |. Do not include any Markdown formatting (no triple backticks, no language tags). Only output the raw Mermaid code.
\end{lstlisting}
\end{promptbox}

\subsubsection{Concept Graph}

The following prompt is used to generate the concept graph using the user's learning content.
\begin{promptbox}
\begin{lstlisting}
Given the following learning content, generate a conceptual diagram of this text.

The graph should meet the following requirements:
- The nodes of the graph represent physical objects presented in the learning content.
- The links represent the relationships connecting them.
- The graph contains relationship that are not necessarily stated in the learning content but can be inferred.
- The graph is as abstract and simple as possible.
- If the graph has math or numbers involved, they are included.
- Each node has a description in square brackets.
- Each link a description that explains its meaning.

NO Explanation.
${mermaidDirections}
 ${learningContent}
\end{lstlisting}
\end{promptbox}

\subsubsection{Scenario Graph} The following prompt is used to generate the scenario graph using the concept graph and the user's chosen scenario.

\begin{promptbox}
\begin{lstlisting}
Given the concept graph and scenario, change the names of all the nodes in the graph to represent the specific objects that the scenario involves. I'm trying to understand how the scenario can represent the relationship that the graph explains. Each node should have an updated is that is specific to this scenario.

A good response:
- Keeps the ids of the nodes in the graph the same (ids being the values in square brackets)
NO Explanation.
${mermaidDirections}
Concept graph:${graph}
Scenario:${scenario}
\end{lstlisting}
\end{promptbox}

\subsubsection{Learning Goal Graph}
The following prompt is used to generate the learning goal graph using the scenario graph and the user's chosen learning goal (labeled hypothesis below).

\begin{promptbox}
\begin{lstlisting}
Given the concept graph and learning goal, give me an updated graph (also using mermaidjs) that only contains nodes and links that pertain to the given learning goal (in other words, remove unnecessary nodes and links).
NO Explanation
${mermaidDirections}
Concept Graph:${graph}
Learning Goal:${hypothesis}
\end{lstlisting}
\end{promptbox}

\subsubsection{User Interaction Graph}

The following prompt is used to generate the user interaction graph using the learning goal graph and the experimental procedure. Subsequent subsections describe how this procedure is generated.
\begin{promptbox}
\begin{lstlisting}
Given an experimental procedure and concept graph, create a UI interaction graph of a web-based interactive simulation for this procedure.
Consider what the main experimental object is in this experiment, along with the dependent and independent variables.
What is the best way to visually show the experimental object within the procedure?
Include nodes for:
- all nodes in the given concept graph that you think are necessary for following the procedure and learning the conclusion of the learning goal. Keep the same ids and labels for these nodes as were in the original concept graph.
- UI controls necessary for following the procedure
Include edges for:
- all edges in the given concept graph that you think are necessary for following the procedure and learning the conclusion of the learning goal.
- any more relationships between visuals
- relationships between the UI controls and the visuals

Give the graph using mermaid js. NO explanation.
${mermaidDirections}

Graph:${graph}
Procedure:${proc}
Learning Goal:${hypothesis}
\end{lstlisting}
\end{promptbox}

\textbf{Descriptive Learning Goals:}
When the selected learning goal is descriptive, procedure is generated by first identifying the independent and dependent variables in question:

\begin{promptbox}
\begin{lstlisting}
In an experiment testing the provided hypothesis using the laws explained in the provided concept graph, what is the main experimental object that we are interacting with in this experiment? NO explanation.

Concept Graph:${graph}
Hypothesis: ${hypothesis}
\end{lstlisting}
\end{promptbox}

\begin{promptbox}
\begin{lstlisting}
In an experiment testing the provided hypothesis using the laws explained in the provided concept graph, what is the dependent variable of this experiment? NO explanation and don't put a period at the end.

Concept Graph:${graph}
Hypothesis: ${hypothesis}
\end{lstlisting}
\end{promptbox}

Using this information, we prompt for a procedure:

\begin{promptbox}
\begin{lstlisting}
You are an expert in designing experimental procedures for interactive simulations. Given the concept graph and learning goal that you are trying to test, generate a simple procedure testing what effect ${indep} has on ${dep} as indicated by the learning goal using the concept graph.
            
A good procedure:
- Comes to the conclusion of the learning goal
- Is simple to follow
- Outlines any data collection that you want to perform
            
NO explanation
            
Concept Graph:${graph}
Learning Goal:${hypothesis}
\end{lstlisting}
\end{promptbox}

\textbf{Explanatory:}
When the selected learning goal is explanatory, procedure is generated by again identifying the independent and dependent variables in question, then prompting for a procedure:

\begin{promptbox}
\begin{lstlisting}
Based on the concept graph, what underlying process explains why ${indep} has an effect on ${dep}? Give me one single phrase. NO explanation and don't put a period at the end.

Concept Graph:${graph}
\end{lstlisting}
\end{promptbox}

\begin{promptbox}
\begin{lstlisting}
You are an expert in designing experimental procedures for interactive simulations. Given the concept graph and learning goal that I am trying to test, give me a simple procedure testing how ${exp} as indicated by the learning goal using the concept graph.

A good procedure:
- Comes to the conclusion of the learning goal
- Is simple to follow
- Outlines any data collection that you want to perform
            
NO explanation.
            
Concept Graph:${graph}
Learning Goal:${hypothesis}
\end{lstlisting}
\end{promptbox}

\textbf{Procedural:}
When the selected learning goal is procedural, we first prompt for the experimental object:

\begin{promptbox}
\begin{lstlisting}
In an experiment testing the provided hypothesis using the laws explained in the provided concept graph, what is the main experimental object that we are interacting with in this experiment? NO explanation.

Concept Graph:${graph}
Hypothesis: ${hypothesis}
\end{lstlisting}
\end{promptbox}

Then we prompt for the process in question:
\begin{promptbox}
\begin{lstlisting}
In an experiment testing the provided hypothesis using the laws explained in the provided concept graph, what is the porocess that the ${obj} goes through? NO explanation and don't put a period at the end.

Concept Graph:${graph}
Hypothesis: ${hypothesis}
\end{lstlisting}
\end{promptbox}

And finally, we use this information to prompt for a procedure:
\begin{promptbox}
\begin{lstlisting}
You are an expert in designing experimental procedures for interactive simulations. Given the concept graph and learning goal that I am trying to test, give me a simple procedure for running an interactive simulation showing how ${obj} goes through ${proc}.

A good procedure:
- Comes to the conclusion of the learning goal
- Is simple to follow
- Outlines any data collection that you want to perform
            
NO explanation.
            
Concept Graph:${graph}
Learning Goal:${hypothesis}
\end{lstlisting}
\end{promptbox}

\subsubsection{Code}

We purposefully want \system simulations to look low-fidelity. In order to achieve this look, we use rough.js. Therefore, any prompts generating code include the following information about rough.js:
\begin{promptbox}
\begin{lstlisting}
Imports and uses roughjs (https://unpkg.com/roughjs@latest/bundled/rough.js) for the entire app, including UI controls. Make sure to include a canvas element in the html to reference as the rough.js canvas in your script.
\end{lstlisting}
\end{promptbox}

The following prompt is used to generate the simulation code:
\begin{promptbox}
\begin{lstlisting}
Create a web-based interactive simulation based on the UI Interface Graph. Do your best to interpret which nodes represent visuals, which represent UI controls & data collection mechanisms, and give each UI control its desired function.
Include:
- A title and descripton of the experiment. The description should act as an introduction for the students to the experiment and what it teaches. In this description include any definitions that directly relate to the learning goal. Don't just outright state the learning goal, we want students to figure this out on their own.
- Integrate any instruction necessary. It should be clear how students should interact with the simulation.
- All nodes representing visuals and phenomena in a visual display of the experiment.
- All nodes representing UI controls & data collection below the visual display. You must label what each UI control represents. If a control represent an amount, make sure to provide concrete experimental units.
- ${roughDirections}

Make sure to include ALL nodes in the UI interface graph and give them their desired purpose
When inplementing the relationship between any two nodes, think about what attributes of each node define the functional reationship between them.
Keep track of these attributes in your script.
        
The simulation is for middle school kids, so make the visuals fun and engaging.
The UI controls should invoke expressive animations in the visual display.
Generate SVGs that are as realistic as possible and use gradients and additional shapes if needed for any necessary experimental objects. Rather than placing SVGs directly in the code, clearly define each as a variable.
For all text, either use the font cabin-sketch-regular or cabin-sketch-bold. In order to use these fonts, add <link rel="preconnect" href="https://fonts.googleapis.com"> <link rel="preconnect" href="https://fonts.gstatic.com" crossorigin> <link href="https://fonts.googleapis.com/css2?family=Cabin+Sketch:wght@400;700&family=Londrina+Sketch&family=Roboto:ital,wght@0,100;0,300;0,400;0,500;0,700;0,900;1,100;1,300;1,400;1,500;1,700;1,900&display=swap" rel="stylesheet"> to the HTML code.

Give the HTML, CSS and any necessary JS interactivity.
Make sure the code displays all visual elements correctly and no uncaught errors occur when it is run.
        
Add comprehensive logging for debugging purposes ONLY. Include the following debug logging system in your code:
    // Debug logging system - only active when window.LOG_DEBUG is true
    function logDebug(message) {
        if (window.LOG_DEBUG) {
            console.log(\`DEBUG [\${new Date().toISOString()}]: \${message}\`);
        }
    }

    // Add this in your initialization
    document.addEventListener('DOMContentLoaded', () => {
        if (window.LOG_DEBUG) {
            logDebug('Simulation initialized');
            // Log all SVG elements
            document.querySelectorAll('svg').forEach((svg, index) => {
                const rect = svg.getBoundingClientRect();
                logDebug(\`SVG #\${index}: Position {x: \${rect.x}, y: \${rect.y}}, Size {w: \${rect.width}, h: \${rect.height}}\`);
            });
            // Log UI controls
            document.querySelectorAll('input, button, select').forEach((control) => {
                logDebug(\`Control \${control.id || 'unnamed'} (\${control.tagName}): Initial value: \${control.value || 'N/A'}\`);
            });
            
            // Add event listeners for logging interactions
            document.querySelectorAll('button').forEach((button) => {
                button.addEventListener('click', () => logDebug(\`Button \${button.id || 'unnamed'} clicked\`));
            });
            document.querySelectorAll('input').forEach((input) => {
                input.addEventListener('change', (e) => logDebug(\`Input \${input.id || 'unnamed'} changed to \${e.target.value}\`));
            });
        }
    });

    // For animations and calculations, add logging in those functions
    // Example: logDebug(\`Calculated \${variable} = \${formula} = \${result}\`);
    // Example: logDebug(\`Element \${id} moved to {x: \${newX}, y: \${newY}}\`);
    
IMPORTANT: This logging should only be active when window.LOG_DEBUG is true, so it won't affect normal usage.
The default state MUST be window.LOG_DEBUG = false; in your code.
Make sure to call logDebug() in key places of your code to track:
- All SVG elements' positions and sizes
- UI control state changes
- User interactions with timestamps
- Animation frame-by-frame updates
- Mathematical calculations with formulas
- Any errors or warnings 

NO explanation.

UI Interface Graph::${graph}
Learning Goal:${hypothesis}
\end{lstlisting}
\end{promptbox}

\subsection{User Direction}
\subsubsection{Scenario Options}

Below is the prompt used to generate potential scenarios given the concept graph.
\begin{promptbox}
\begin{lstlisting}
Give me 8 contexts that I can use to teach the concepts involved in the concept graph.
By context, I mean an instantiation of the concepts involved in the concept graph.
For each context, write a title enclosed in double curly braces {{title}} without numbering and a few words on how it presents the concept and how the teacher should use the contexts to teach high school kids about this concept.
Don't talk about classroom activities or student activities.
MAKE SURE to separate each context with "|".
            
Concept graph:${graph}
\end{lstlisting}
\end{promptbox}

\subsubsection{Learning Goal Options}
Below is the prompt used to generate potential learning goals given the scenario graph.

\begin{promptbox}
\begin{lstlisting}
Give me 6 concise and testable learning goals that I can use to teach high schoolers about the concept involved in the concept map. When forming goals, please focus on a few specific nodes in the concept graph and how the links between them represent relationships. Phrase these goals as a single statement that you want the student to learn by looking at the concept map, sort of like a hypothesis. For each learning goal, write the goal in double curly braces {{title}} without numbering and a brief description of what learning gains that goal will lead to. If the goal just describes a process, begin the desciption with "1.", if the goal explains why a process works in the way it does, begin the description with "2.", and if the goal explains an overall process, begin the description with "3.". MAKE SURE to separate each learning goal from the next with "|".

Concept Graph: ${graph}
\end{lstlisting}
\end{promptbox}

\subsection{Abstraction Modification}

\subsubsection{Suggesting a Code Change}
The following prompt is used to suggest a class of code change based on a chat or error description message.

\begin{promptbox}
\begin{lstlisting}
In a previous prompt, you generated the provided HTML code for an interactive simulation based on the provided UI Map.

However, there's an issue with this code/UI map: ${prompt}.

This issue may also be outlined via the red annotations in the provided image, which are labeled with labels such as "A1" or "A2" near the annotation.
These labels may be referenced in the above issue description.
        
Given the issue mentioned above, what type of change to the code would you would need to make in order to fix this issue? Choose from the following types of changes:
1. Add new variable/visual component to the code.
2. Add new function that acts on/using specific variables/visual components to the code.
3. Remove a variable/visual component from the code.
4. Remove a function that acts on/using specific variables/visual components from the code.
5. Completely re-implement a variable/visual component in the code.
6. Completely re-implement a function in the code.
7. Change an SVG in the code.
8. Change the implementation of some preexisting functionality.

Given me just the number associated with the type of change you would make.
NO EXPLAINATION.
\end{lstlisting}
\end{promptbox}

\subsubsection{Populate Remove Edge}
If the chat suggests removing an edge, we populate a chat widget describing the change to the user.
The following prompt is used to get the necessary information.
\begin{promptbox}
\begin{lstlisting}
In a previous prompt, you generated HTML code for an interactive simulation based on the provided UI Map.

However, there's an issue with this code/UI map: ${prompt}.

This issue may also be outlined via the red annotations in the provided image, which are labeled with labels such as "A1" or "A2" near the annotation.
These labels may be referenced in the above issue description.
        
We think removing an edge from the UI Map might solve the above issue conceptually.
We want some more information on the edge that should be removed.
 
Please respond in the following format: 
{ type: 8, message: <SHORT message explaining your suggestion. Use full, eloquent sentences and first person.>, source: <id of the source node>, target: <id of the target node>, label: <the label of the edge to delete>}

NO EXPLAINATION.
\end{lstlisting}
\end{promptbox}

\subsubsection{Populate Remove Node}
\begin{promptbox}
\begin{lstlisting}
In a previous prompt, you generated HTML code for an interactive simulation based on the provided UI Map.

However, there's an issue with this code/UI map: ${prompt}.

This issue may also be outlined via the red annotations in the provided image, which are labeled with labels such as "A1" or "A2" near the annotation.
These labels may be referenced in the above issue description.
        
We think removing a node from the UI Map might solve the above issue conceptually.
We want some more information on the node that should be removed.

Please respond in the following format: { type: 7, message: <SHORT message explaining your suggestion. Use full, eloquent sentences and first person.>, node: <id of the node to delete>}

NO EXPLAINATION.
\end{lstlisting}
\end{promptbox}

\subsubsection{Populate Edit Edge}
\begin{promptbox}
\begin{lstlisting}
In a previous prompt, you generated HTML code for an interactive simulation based on the provided UI Map.

However, there's an issue with this code/UI map: ${prompt}.

This issue may also be outlined via the red annotations in the provided image, which are labeled with labels such as "A1" or "A2" near the annotation.
These labels may be referenced in the above issue description.
        
We think changing the label of an edge in the UI Map might solve the above issue conceptually.
We want some more information on the edge that should be changed.

Please respond in the following format: 
{ type: 6, message: <SHORT message explaining your suggestion. Use full, eloquent sentences and first person.>, source: <id of the source node>, target: <id of the target node>, oldLabel: <the old label of the edge to change>, newLabel: <the new label>}

NO EXPLAINATION.
\end{lstlisting}
\end{promptbox}

\subsubsection{Populate Edit Node}
\begin{promptbox}
\begin{lstlisting}
In a previous prompt, you generated HTML code for an interactive simulation based on the provided UI Map.

However, there's an issue with this code/UI map: ${prompt}.

This issue may also be outlined via the red annotations in the provided image, which are labeled with labels such as "A1" or "A2" near the annotation.
These labels may be referenced in the above issue description.
        
We think changing the label of a node in the UI Map might solve the above issue conceptually.
We want some more information on the node that should be changed.

Please respond in the following format:
{ type: 5, message: <SHORT message explaining your suggestion. Use full, eloquent sentences and first person.>, node: <id of the node to change>, oldLabel: <the old label of that node>, newLabel: <the new label>}
        
NO EXPLAINATION.
\end{lstlisting}
\end{promptbox}

\subsubsection{Populate Redraw}
\begin{promptbox}
\begin{lstlisting}
In a previous prompt, you generated the provided HTML code for an interactive simulation.

However, there's an issue with this code/UI map: ${prompt}.

This issue may also be outlined via the red annotations in the provided image, which are labeled with labels such as "A1" or "A2" near the annotation.
These labels may be referenced in the above issue description.
        
We think editing one of the SVGs in the code might solve the above issue.
We want some more information on how we should update the SVG.

Please respond in the following format:
{ type: 4, message: <SHORT message explaining your suggestion. Use full, eloquent sentences and first person.>, box: <a box surrounding the object that is being redrawn in the provided image. First consider which object needs to be redrawn. Then, find the EXACT location and size of this object by looking at both the provided image and the html code. Based on this location and size, generate a box that surrounds this object using the format [start_x, start_y, width, height] that uses pixels as its units (e.g. [0,0,50,50] would be a 50x50px box in the top left corner of the image). When applied on top of the image, the box should completely encompass the object.>, svg: <simple svg representing the new visual, approx 100px by 100px> }
        
NO EXPLAINATION.
\end{lstlisting}
\end{promptbox}

\subsubsection{Populate Edit Assumptions}
\begin{promptbox}
\begin{lstlisting}
In a previous prompt, you generated HTML code for an interactive simulation based on the provided UI Map.

However, there's an issue with this code/UI map: ${prompt}.

This issue may also be outlined via the red annotations in the provided image, which are labeled with labels such as "A1" or "A2" near the annotation.
These labels may be referenced in the above issue description.
        
We think editing the details of one of the nodes in the UI Map might solve the above issue.
We want some more information on how we should edit these details to fix the issue.

Please respond in the following format:
{ type: 3, message: <SHORT message explaining your suggestion. Use full, eloquent sentences and first person.>, node: <id of the node whose assumptions are being updated>, assumptions: <list of assumptions updated to explicitly fix the issue> }

NO EXPLAINATION.
\end{lstlisting}
\end{promptbox}

\subsubsection{Populate Add Edge}
\begin{promptbox}
\begin{lstlisting}
In a previous prompt, you generated HTML code for an interactive simulation based on the provided UI Map.

However, there's an issue with this code/UI map: ${prompt}.

This issue may also be outlined via the red annotations in the provided image, which are labeled with labels such as "A1" or "A2" near the annotation.
These labels may be referenced in the above issue description.
        
We think adding an edge to the UI Map might solve the above issue conceptually.
We want some more information on the edge that should be added.
   
Please respond in the following format: 
{ type: 2, message: <SHORT message explaining your suggestion. Use full, eloquent sentences and first person.>, source: <id of the source node>, target: <id of the target node>, label: <new edge name>}

NO EXPLAINATION.
\end{lstlisting}
\end{promptbox}

\subsubsection{Populate Add Node}
\begin{promptbox}
\begin{lstlisting}
In a previous prompt, you generated HTML code for an interactive simulation based on the provided UI Map.

However, there's an issue with this code/UI map: ${prompt}.

This issue may also be outlined via the red annotations in the provided image, which are labeled with labels such as "A1" or "A2" near the annotation.
These labels may be referenced in the above issue description.
        
We think adding a node to the UI Map might solve the above issue conceptually.
We want some more information on the node that should be added.

Please respond in the following format:
{ type: 1, message: <SHORT message explaining your suggestion. Use full, eloquent sentences and first person.>, label: <new node name> }

NO EXPLAINATION.
\end{lstlisting}
\end{promptbox}

\begin{promptbox}
\begin{lstlisting}
Your designers want to know what nodes and links in the given UI interactivity graph correspond to the region(s) circled in red in your prototype.
Given an image, the code used to generate the prototype in the image, and a graph, you respond with the subgraph corresponding to the elements and relationships these region(s) represents.
        
A good subgraph has the:
- nodes corresponding to the visuals/elements circled in red
- nodes corresponding to the attributes of the elements circled in red
- links corresponding to the relationships between all included nodes
- original label for EVERY node and edge
        
NO explanation. Only respond with the mermaidjs graph.
\end{lstlisting}
\end{promptbox}

\subsubsection{Code Assumptions}
The following prompt is used to indentify the current code assumptions abstraction.
\begin{promptbox}
\begin{lstlisting}
Given the following html code and the UI Interactivity Map that it represents, please give me a json object with an attribute for each node in the UI Map . The name of the each attribute should be the label, or value in square brackets, of each node (DO NOT NAME IT THE NAME IN SQUARE BRACKETS). The value for each of these attributes should be a list of all of the details and assumptions that the code is making about this node. For example, the numeric range of a node representing a slider, attributes of a phypical display or graph, all formulas used to caluculate displayed or output values, and the way physical objects are visually represented.

A good list of assumptions:
- Doesn't reference specific variables in the code but instead speaks generically
- Pays close attention to the implementation in order to identify assumptions that are not intended (i.e. bugs)
            
NO explaination.
            
Html Code: ${htmlCode}
UI Map: ${UIMap}
\end{lstlisting}
\end{promptbox}

The following prompt is used to update the code based on the user's changes to the code assumptions.
\begin{promptbox}
\begin{lstlisting}
The given html code is the implementation of an experiment based on the following concept graph. Please update the html code so that it implements the details outlined in the given details list for the ${node} object.

A good response:
- contains all the same functionality of the original html code
- updates the code so that ${node} now abides by the provided list of details
- ${roughDirections}

NO explanation.

Graph:${graph}
Html Code:${htmlCode}
Node of interest: ${node}
Details:${JSON.stringify(newAssumptions)}
\end{lstlisting}
\end{promptbox}

\textbf{Redraw}

The following prompt is used to redraw a visual when the LLM identifies that the user wants to update it (from a chat or error description message).
\begin{promptbox}
\begin{lstlisting}
Given the above low-fidelity hand-drawn image, create a high fidelity SVG representing the object that the hand-drawn image is of.

A good SVG response:
- interprets the object that the hand-drawn image is of (typically this object is one of the objects included as nodes in the provided UI Map)
- has similar shape and features to the hand drawn image
- uses gradients and additional shapes if needed
- ignores the fact that the hand-drawn sketch is drawn in red

NO EXPLAINATION
\end{lstlisting}
\end{promptbox}

The following prompt updated the simulation code to use a new svg for a specific visual.
\begin{promptbox}
\begin{lstlisting}
Given the above image and html code, replace the SVG circled in red in the image with the provided SVG in the code. Respond with the ENTIRE updated HTML code and nothing selse.

A good response MUST:
- have a the old circled SVG replaced with the new provided SVG
- resize the new SVG so it is the same size as the original (not just cropping the new SVG, but actaully reducing/increasing its attributes)
- have all of the functionality of the original code
- ${roughDirections}

NO EXPLAINATION
\end{lstlisting}
\end{promptbox}

\subsubsection{Auto-Add Edges}
Below is the prompt used to automatically add links when a new node is added by the user.
\begin{promptbox}
\begin{lstlisting}
Please add a new node to the following mermaid graph with the provided label. Add any edge that you think reasonably connect this new object to the rest of the graph.

NO explaination.
${mermaidDirections}

Graph: ${UIMap}
Label: ${newNodeName}
\end{lstlisting}
\end{promptbox}

\subsection{Automated Testing}

The following prompt is used to generate test cases for automated testing.
\begin{promptbox}
\begin{lstlisting}
You had previously created the following HTML code based on the UI Map and learning goal.
More info:
- HTML Code: ${htmlCode}
- Learning Goal: ${selectedHypothesis}
- UI Map: ${UIMap}.
Please generate a JSON array of structured test cases for the interactive simulation.
Each test case should be a JSON object with the following keys:
- uiElementId: The ID of the UI element to interact with.
- actionType: The type of action (one of "click", "set_value", "toggle", "verify_content").
- actionValue (optional): The value to set (if applicable).
- description: A brief description of what is being tested for.
- expectedOutcome: A brief description of the expected result.
- isUIVerification: A boolean indicating whether this test case is related to changes in any of the UI components or visual elements in the simulation.
Return only the JSON array, with no additional text, and ensure that the JSON starts between <START> and <STOP> tags.
\end{lstlisting}
\end{promptbox}

Below is the prompt used to verify test results using debug log information and Puppeteer screenshots.
\begin{promptbox}
\begin{lstlisting}
You had previously created the following HTML code based on the UI Map and learning goal.
More info:
- HTML Code: ${htmlCode}
- UI Map: ${UIMap}
- Learning Goal: ${selectedHypothesis}
- JavaScript Errors captured: ${.join('; ')}.
${debugLogsText}
${initialScreenshotNote}
Note: An initial screenshot of the UI (before any interactions) is available.

Based on the above context, including the test case results, screenshots and runtime logs, do the following:
1) Verify whether the actual outcomes in each test case (as described below) match the expected outcomes.
2) If discrepancies or errors remain, update the HTML code so that all test cases pass and errors are resolved.
3) Verify that the simulation satisfies the learning goal.
IMPORTANT: Return only the updated HTML code, starting between <START> and <STOP> tags, or return "PASS" if no changes are needed.

Here is the:
1) Structured list of test case results and verification details,
2) Indication of whether the HTML code needs to be changed.
If no changes are needed, simply return "PASS".
If changes are required, return the updated HTML code starting between <START> and <STOP> tags.
\end{lstlisting}
\end{promptbox}

The following prompt is used to fix js errors in the simulation code when they are present.
\begin{promptbox}
\begin{lstlisting}
You had previously created the following HTML code based on the UI Map and learning goal.
More info:
- HTML Code: ${htmlCode}
- UI Map: ${UIMap}.
However, the HTML code is generating these JavaScript errors: ${errorMessages.join('; ')}.
Please update the HTML code so that these errors are fixed while preserving all the original functionality.
Return only the updated HTML code, starting between <START> and <STOP> tags.
\end{lstlisting}
\end{promptbox}

%% file: 00_main.bbl

\begin{thebibliography}{68}

%%% the Web address.

\ifx \showCODEN    \undefined \def \showCODEN     #1{\unskip}     \fi
\ifx \showDOI      \undefined \def \showDOI       #1{#1}\fi
\ifx \showISBNx    \undefined \def \showISBNx     #1{\unskip}     \fi
\ifx \showISBNxiii \undefined \def \showISBNxiii  #1{\unskip}     \fi
\ifx \showISSN     \undefined \def \showISSN      #1{\unskip}     \fi
\ifx \showLCCN     \undefined \def \showLCCN      #1{\unskip}     \fi
\ifx \shownote     \undefined \def \shownote      #1{#1}          \fi
\ifx \showarticletitle \undefined \def \showarticletitle #1{#1}   \fi
\ifx \showURL      \undefined \def \showURL       {\relax}        \fi
% invisible to TeX
\providecommand\bibfield[2]{#2}
\providecommand\bibinfo[2]{#2}
\providecommand\natexlab[1]{#1}
\providecommand\showeprint[2][]{arXiv:#2}

\bibitem[Adams(2014)]%
        {adams2014fundamentals}
\bibfield{author}{\bibinfo{person}{Ernest Adams}.} \bibinfo{year}{2014}\natexlab{}.
\newblock \bibinfo{booktitle}{\emph{Fundamentals of game design}}.
\newblock \bibinfo{publisher}{Pearson Education}.
\newblock


\bibitem[Agrawal et~al\mbox{.}(2022)]%
        {agrawal2022building}
\bibfield{author}{\bibinfo{person}{Garima Agrawal}, \bibinfo{person}{Yuli Deng}, \bibinfo{person}{Jongchan Park}, \bibinfo{person}{Huan Liu}, {and} \bibinfo{person}{Ying-Chih Chen}.} \bibinfo{year}{2022}\natexlab{}.
\newblock \showarticletitle{Building knowledge graphs from unstructured texts: Applications and impact analyses in cybersecurity education}.
\newblock \bibinfo{journal}{\emph{Information}} \bibinfo{volume}{13}, \bibinfo{number}{11} (\bibinfo{year}{2022}), \bibinfo{pages}{526}.
\newblock


\bibitem[Alaboudi and Latoza(2023)]%
        {alaboudi2023hypothesizer}
\bibfield{author}{\bibinfo{person}{Abdulaziz Alaboudi} {and} \bibinfo{person}{Thomas~D Latoza}.} \bibinfo{year}{2023}\natexlab{}.
\newblock \showarticletitle{Hypothesizer: A Hypothesis-Based Debugger to Find and Test Debugging Hypotheses}. In \bibinfo{booktitle}{\emph{Proceedings of the 36th Annual ACM Symposium on User Interface Software and Technology}}. \bibinfo{pages}{1--14}.
\newblock


\bibitem[Angert et~al\mbox{.}(2023)]%
        {angert2023spellburst}
\bibfield{author}{\bibinfo{person}{Tyler Angert}, \bibinfo{person}{Miroslav Suzara}, \bibinfo{person}{Jenny Han}, \bibinfo{person}{Christopher Pondoc}, {and} \bibinfo{person}{Hariharan Subramonyam}.} \bibinfo{year}{2023}\natexlab{}.
\newblock \showarticletitle{Spellburst: A node-based interface for exploratory creative coding with natural language prompts}. In \bibinfo{booktitle}{\emph{Proceedings of the 36th Annual ACM Symposium on User Interface Software and Technology}}. \bibinfo{pages}{1--22}.
\newblock


\bibitem[Anthropic(2023)]%
        {anthropic2023claude}
\bibfield{author}{\bibinfo{person}{Anthropic}.} \bibinfo{year}{2023}\natexlab{}.
\newblock \bibinfo{title}{Claude: Language Model by Anthropic}.
\newblock \bibinfo{howpublished}{\url{https://www.anthropic.com/}}.
\newblock


\bibitem[Armeni et~al\mbox{.}(2019)]%
        {Armeni20193DSG}
\bibfield{author}{\bibinfo{person}{Iro Armeni}, \bibinfo{person}{Zhi-Yang He}, \bibinfo{person}{JunYoung Gwak}, \bibinfo{person}{Amir Zamir}, \bibinfo{person}{Martin Fischer}, \bibinfo{person}{Jitendra Malik}, {and} \bibinfo{person}{Silvio Savarese}.} \bibinfo{year}{2019}\natexlab{}.
\newblock \showarticletitle{3D Scene Graph: A Structure for Unified Semantics, 3D Space, and Camera}.
\newblock \bibinfo{journal}{\emph{2019 IEEE/CVF International Conference on Computer Vision (ICCV)}} (\bibinfo{year}{2019}), \bibinfo{pages}{5663--5672}.
\newblock
\urldef\tempurl%
\url{https://api.semanticscholar.org/CorpusID:203837042}
\showURL{%
\tempurl}


\bibitem[Austin et~al\mbox{.}(2021)]%
        {austin2021program}
\bibfield{author}{\bibinfo{person}{Jacob Austin}, \bibinfo{person}{Augustus Odena}, \bibinfo{person}{Maxwell Nye}, \bibinfo{person}{Maarten Bosma}, \bibinfo{person}{Henryk Michalewski}, \bibinfo{person}{David Dohan}, \bibinfo{person}{Ellen Jiang}, \bibinfo{person}{Carrie Cai}, \bibinfo{person}{Michael Terry}, \bibinfo{person}{Quoc Le}, {et~al\mbox{.}}} \bibinfo{year}{2021}\natexlab{}.
\newblock \showarticletitle{Program synthesis with large language models}.
\newblock \bibinfo{journal}{\emph{arXiv preprint arXiv:2108.07732}} (\bibinfo{year}{2021}).
\newblock


\bibitem[Barnett and Ruhsam(1968)]%
        {barnett1968natural}
\bibfield{author}{\bibinfo{person}{Michael~P Barnett} {and} \bibinfo{person}{WM Ruhsam}.} \bibinfo{year}{1968}\natexlab{}.
\newblock \showarticletitle{A natural language programming system for text processing}.
\newblock \bibinfo{journal}{\emph{IEEE transactions on engineering writing and speech}} \bibinfo{volume}{11}, \bibinfo{number}{2} (\bibinfo{year}{1968}), \bibinfo{pages}{45--52}.
\newblock


\bibitem[Bostock and Heer(2009)]%
        {bostock2009protovis}
\bibfield{author}{\bibinfo{person}{Michael Bostock} {and} \bibinfo{person}{Jeffrey Heer}.} \bibinfo{year}{2009}\natexlab{}.
\newblock \showarticletitle{Protovis: A graphical toolkit for visualization}.
\newblock \bibinfo{journal}{\emph{IEEE transactions on visualization and computer graphics}} \bibinfo{volume}{15}, \bibinfo{number}{6} (\bibinfo{year}{2009}), \bibinfo{pages}{1121--1128}.
\newblock


\bibitem[Bouzenia et~al\mbox{.}(2024)]%
        {bouzenia2024repairagent}
\bibfield{author}{\bibinfo{person}{Islem Bouzenia}, \bibinfo{person}{Premkumar Devanbu}, {and} \bibinfo{person}{Michael Pradel}.} \bibinfo{year}{2024}\natexlab{}.
\newblock \showarticletitle{Repairagent: An autonomous, llm-based agent for program repair}.
\newblock \bibinfo{journal}{\emph{arXiv preprint arXiv:2403.17134}} (\bibinfo{year}{2024}).
\newblock


\bibitem[Bradley et~al\mbox{.}(2018)]%
        {bradley2018context}
\bibfield{author}{\bibinfo{person}{Nick~C Bradley}, \bibinfo{person}{Thomas Fritz}, {and} \bibinfo{person}{Reid Holmes}.} \bibinfo{year}{2018}\natexlab{}.
\newblock \showarticletitle{Context-aware conversational developer assistants}. In \bibinfo{booktitle}{\emph{Proceedings of the 40th International Conference on Software Engineering}}. \bibinfo{pages}{993--1003}.
\newblock


\bibitem[Brady et~al\mbox{.}(2022)]%
        {brady2022block}
\bibfield{author}{\bibinfo{person}{Corey Brady}, \bibinfo{person}{Brian Broll}, \bibinfo{person}{Gordon Stein}, \bibinfo{person}{Devin Jean}, \bibinfo{person}{Shuchi Grover}, \bibinfo{person}{Veronica Catet{\'e}}, \bibinfo{person}{Tiffany Barnes}, {and} \bibinfo{person}{{\'A}kos L{\'e}deczi}.} \bibinfo{year}{2022}\natexlab{}.
\newblock \showarticletitle{Block-based abstractions and expansive services to make advanced computing concepts accessible to novices}.
\newblock \bibinfo{journal}{\emph{Journal of Computer Languages}}  \bibinfo{volume}{73} (\bibinfo{year}{2022}), \bibinfo{pages}{101156}.
\newblock


\bibitem[Cheema and LaViola(2012)]%
        {cheema2012physicsbook}
\bibfield{author}{\bibinfo{person}{Salman Cheema} {and} \bibinfo{person}{Joseph LaViola}.} \bibinfo{year}{2012}\natexlab{}.
\newblock \showarticletitle{PhysicsBook: a sketch-based interface for animating physics diagrams}. In \bibinfo{booktitle}{\emph{Proceedings of the 2012 ACM international conference on Intelligent User Interfaces}}. \bibinfo{pages}{51--60}.
\newblock


\bibitem[Chen et~al\mbox{.}(2018)]%
        {chen2018knowedu}
\bibfield{author}{\bibinfo{person}{Penghe Chen}, \bibinfo{person}{Yu Lu}, \bibinfo{person}{Vincent~W Zheng}, \bibinfo{person}{Xiyang Chen}, {and} \bibinfo{person}{Boda Yang}.} \bibinfo{year}{2018}\natexlab{}.
\newblock \showarticletitle{Knowedu: A system to construct knowledge graph for education}.
\newblock \bibinfo{journal}{\emph{Ieee Access}}  \bibinfo{volume}{6} (\bibinfo{year}{2018}), \bibinfo{pages}{31553--31563}.
\newblock


\bibitem[Chopra et~al\mbox{.}(2024)]%
        {chopra2024exploring}
\bibfield{author}{\bibinfo{person}{Bhavya Chopra}, \bibinfo{person}{Yasharth Bajpai}, \bibinfo{person}{Param Biyani}, \bibinfo{person}{Gustavo Soares}, \bibinfo{person}{Arjun Radhakrishna}, \bibinfo{person}{Chris Parnin}, {and} \bibinfo{person}{Sumit Gulwani}.} \bibinfo{year}{2024}\natexlab{}.
\newblock \showarticletitle{Exploring Interaction Patterns for Debugging: Enhancing Conversational Capabilities of AI-assistants}.
\newblock \bibinfo{journal}{\emph{arXiv preprint arXiv:2402.06229}} (\bibinfo{year}{2024}).
\newblock


\bibitem[Denny et~al\mbox{.}(2023)]%
        {10.1145/3545945.3569823}
\bibfield{author}{\bibinfo{person}{Paul Denny}, \bibinfo{person}{Viraj Kumar}, {and} \bibinfo{person}{Nasser Giacaman}.} \bibinfo{year}{2023}\natexlab{}.
\newblock \showarticletitle{Conversing with Copilot: Exploring Prompt Engineering for Solving CS1 Problems Using Natural Language}. In \bibinfo{booktitle}{\emph{Proceedings of the 54th ACM Technical Symposium on Computer Science Education V. 1}} (Toronto ON, Canada) \emph{(\bibinfo{series}{SIGCSE 2023})}. \bibinfo{publisher}{Association for Computing Machinery}, \bibinfo{address}{New York, NY, USA}, \bibinfo{pages}{1136–1142}.
\newblock
\showISBNx{9781450394314}
\urldef\tempurl%
\url{https://doi.org/10.1145/3545945.3569823}
\showDOI{\tempurl}


\bibitem[Dhuliawala et~al\mbox{.}(2023)]%
        {dhuliawala2023chain}
\bibfield{author}{\bibinfo{person}{Shehzaad Dhuliawala}, \bibinfo{person}{Mojtaba Komeili}, \bibinfo{person}{Jing Xu}, \bibinfo{person}{Roberta Raileanu}, \bibinfo{person}{Xian Li}, \bibinfo{person}{Asli Celikyilmaz}, {and} \bibinfo{person}{Jason Weston}.} \bibinfo{year}{2023}\natexlab{}.
\newblock \showarticletitle{Chain-of-verification reduces hallucination in large language models}.
\newblock \bibinfo{journal}{\emph{arXiv preprint arXiv:2309.11495}} (\bibinfo{year}{2023}).
\newblock


\bibitem[Dou et~al\mbox{.}(2024)]%
        {dou2024s}
\bibfield{author}{\bibinfo{person}{Shihan Dou}, \bibinfo{person}{Haoxiang Jia}, \bibinfo{person}{Shenxi Wu}, \bibinfo{person}{Huiyuan Zheng}, \bibinfo{person}{Weikang Zhou}, \bibinfo{person}{Muling Wu}, \bibinfo{person}{Mingxu Chai}, \bibinfo{person}{Jessica Fan}, \bibinfo{person}{Caishuang Huang}, \bibinfo{person}{Yunbo Tao}, {et~al\mbox{.}}} \bibinfo{year}{2024}\natexlab{}.
\newblock \showarticletitle{What's Wrong with Your Code Generated by Large Language Models? An Extensive Study}.
\newblock \bibinfo{journal}{\emph{arXiv preprint arXiv:2407.06153}} (\bibinfo{year}{2024}).
\newblock


\bibitem[Durman(2024)]%
        {joint}
\bibfield{author}{\bibinfo{person}{David Durman}.} \bibinfo{year}{2024}\natexlab{}.
\newblock \bibinfo{title}{joint.js}.
\newblock \bibinfo{howpublished}{\url{https://www.jointjs.com/}}.
\newblock


\bibitem[Fan et~al\mbox{.}(2023)]%
        {fan2023automated}
\bibfield{author}{\bibinfo{person}{Zhiyu Fan}, \bibinfo{person}{Xiang Gao}, \bibinfo{person}{Martin Mirchev}, \bibinfo{person}{Abhik Roychoudhury}, {and} \bibinfo{person}{Shin~Hwei Tan}.} \bibinfo{year}{2023}\natexlab{}.
\newblock \showarticletitle{Automated repair of programs from large language models}. In \bibinfo{booktitle}{\emph{2023 IEEE/ACM 45th International Conference on Software Engineering (ICSE)}}. IEEE, \bibinfo{pages}{1469--1481}.
\newblock


\bibitem[Foundation(2024)]%
        {node}
\bibfield{author}{\bibinfo{person}{OpenJS Foundation}.} \bibinfo{year}{2024}\natexlab{}.
\newblock \bibinfo{title}{Node.js}.
\newblock \bibinfo{howpublished}{\url{https://nodejs.org/en/}}.
\newblock


\bibitem[Gao et~al\mbox{.}(2024)]%
        {gao2024efficient}
\bibfield{author}{\bibinfo{person}{Silin Gao}, \bibinfo{person}{Jane Dwivedi-Yu}, \bibinfo{person}{Ping Yu}, \bibinfo{person}{Xiaoqing~Ellen Tan}, \bibinfo{person}{Ramakanth Pasunuru}, \bibinfo{person}{Olga Golovneva}, \bibinfo{person}{Koustuv Sinha}, \bibinfo{person}{Asli Celikyilmaz}, \bibinfo{person}{Antoine Bosselut}, {and} \bibinfo{person}{Tianlu Wang}.} \bibinfo{year}{2024}\natexlab{}.
\newblock \showarticletitle{Efficient tool use with chain-of-abstraction reasoning}.
\newblock \bibinfo{journal}{\emph{arXiv preprint arXiv:2401.17464}} (\bibinfo{year}{2024}).
\newblock


\bibitem[Google(2024)]%
        {firebase}
\bibfield{author}{\bibinfo{person}{Google}.} \bibinfo{year}{2024}\natexlab{}.
\newblock \bibinfo{title}{Firebase}.
\newblock \bibinfo{howpublished}{\url{https://firebase.google.com/}}.
\newblock


\bibitem[Green(1989)]%
        {green1989cognitive}
\bibfield{author}{\bibinfo{person}{Thomas~RG Green}.} \bibinfo{year}{1989}\natexlab{}.
\newblock \showarticletitle{Cognitive dimensions of notations}.
\newblock \bibinfo{journal}{\emph{People and computers V}} (\bibinfo{year}{1989}), \bibinfo{pages}{443--460}.
\newblock


\bibitem[Hart(1988)]%
        {hart1988development}
\bibfield{author}{\bibinfo{person}{SG Hart}.} \bibinfo{year}{1988}\natexlab{}.
\newblock \showarticletitle{Development of NASA-TLX (Task Load Index): Results of empirical and theoretical research}.
\newblock \bibinfo{journal}{\emph{Human mental workload/Elsevier}} (\bibinfo{year}{1988}).
\newblock


\bibitem[Herskovitz et~al\mbox{.}(2024)]%
        {Herskovitz2024ProgramAllyCC}
\bibfield{author}{\bibinfo{person}{Jaylin Herskovitz}, \bibinfo{person}{Andi Xu}, \bibinfo{person}{Rahaf Alharbi}, {and} \bibinfo{person}{Anhong Guo}.} \bibinfo{year}{2024}\natexlab{}.
\newblock \showarticletitle{ProgramAlly: Creating Custom Visual Access Programs via Multi-Modal End-User Programming}.
\newblock \bibinfo{journal}{\emph{ArXiv}}  \bibinfo{volume}{abs/2408.10499} (\bibinfo{year}{2024}).
\newblock
\urldef\tempurl%
\url{https://api.semanticscholar.org/CorpusID:271909496}
\showURL{%
\tempurl}


\bibitem[Holowaychuk(2024)]%
        {express}
\bibfield{author}{\bibinfo{person}{TJ Holowaychuk}.} \bibinfo{year}{2024}\natexlab{}.
\newblock \bibinfo{title}{Express.js}.
\newblock \bibinfo{howpublished}{\url{https://expressjs.com/}}.
\newblock


\bibitem[Hu et~al\mbox{.}(2024)]%
        {hu2024deploying}
\bibfield{author}{\bibinfo{person}{Zichao Hu}, \bibinfo{person}{Francesca Lucchetti}, \bibinfo{person}{Claire Schlesinger}, \bibinfo{person}{Yash Saxena}, \bibinfo{person}{Anders Freeman}, \bibinfo{person}{Sadanand Modak}, \bibinfo{person}{Arjun Guha}, {and} \bibinfo{person}{Joydeep Biswas}.} \bibinfo{year}{2024}\natexlab{}.
\newblock \showarticletitle{Deploying and evaluating llms to program service mobile robots}.
\newblock \bibinfo{journal}{\emph{IEEE Robotics and Automation Letters}} \bibinfo{volume}{9}, \bibinfo{number}{3} (\bibinfo{year}{2024}), \bibinfo{pages}{2853--2860}.
\newblock


\bibitem[Hutchins(1995)]%
        {hutchins1995cognition}
\bibfield{author}{\bibinfo{person}{Edwin Hutchins}.} \bibinfo{year}{1995}\natexlab{}.
\newblock \bibinfo{booktitle}{\emph{Cognition in the Wild}}.
\newblock \bibinfo{publisher}{MIT press}.
\newblock


\bibitem[Jiang et~al\mbox{.}(2022a)]%
        {Jiang2022DiscoveringTSA}
\bibfield{author}{\bibinfo{person}{Ellen Jiang}, \bibinfo{person}{Edwin Toh}, \bibinfo{person}{A. Molina}, \bibinfo{person}{Kristen Olson}, \bibinfo{person}{Claire Kayacik}, \bibinfo{person}{Aaron Donsbach}, \bibinfo{person}{Carrie~J. Cai}, {and} \bibinfo{person}{Michael Terry}.} \bibinfo{year}{2022}\natexlab{a}.
\newblock \showarticletitle{Discovering the Syntax and Strategies of Natural Language Programming with Generative Language Models}.
\newblock \bibinfo{journal}{\emph{Proceedings of the 2022 CHI Conference on Human Factors in Computing Systems}} (\bibinfo{year}{2022}).
\newblock
\urldef\tempurl%
\url{http://dl.acm.org/citation.cfm?id=3501870}
\showURL{%
\tempurl}


\bibitem[Jiang et~al\mbox{.}(2022b)]%
        {jiang2022discovering}
\bibfield{author}{\bibinfo{person}{Ellen Jiang}, \bibinfo{person}{Edwin Toh}, \bibinfo{person}{Alejandra Molina}, \bibinfo{person}{Kristen Olson}, \bibinfo{person}{Claire Kayacik}, \bibinfo{person}{Aaron Donsbach}, \bibinfo{person}{Carrie~J Cai}, {and} \bibinfo{person}{Michael Terry}.} \bibinfo{year}{2022}\natexlab{b}.
\newblock \showarticletitle{Discovering the syntax and strategies of natural language programming with generative language models}. In \bibinfo{booktitle}{\emph{Proceedings of the 2022 CHI Conference on Human Factors in Computing Systems}}. \bibinfo{pages}{1--19}.
\newblock


\bibitem[Kazi et~al\mbox{.}(2014)]%
        {kazi2014kitty}
\bibfield{author}{\bibinfo{person}{Rubaiat~Habib Kazi}, \bibinfo{person}{Fanny Chevalier}, \bibinfo{person}{Tovi Grossman}, {and} \bibinfo{person}{George Fitzmaurice}.} \bibinfo{year}{2014}\natexlab{}.
\newblock \showarticletitle{Kitty: sketching dynamic and interactive illustrations}. In \bibinfo{booktitle}{\emph{Proceedings of the 27th annual ACM symposium on User interface software and technology}}. \bibinfo{pages}{395--405}.
\newblock


\bibitem[Ko et~al\mbox{.}(2011)]%
        {Ko2011TheSOA}
\bibfield{author}{\bibinfo{person}{Amy~J. Ko}, \bibinfo{person}{Robin Abraham}, \bibinfo{person}{Laura Beckwith}, \bibinfo{person}{Alan~F. Blackwell}, \bibinfo{person}{Margaret~M. Burnett}, \bibinfo{person}{Martin Erwig}, \bibinfo{person}{Chris Scaffidi}, \bibinfo{person}{Joseph Lawrance}, \bibinfo{person}{Henry Lieberman}, \bibinfo{person}{Brad~A. Myers}, \bibinfo{person}{M. Rosson}, \bibinfo{person}{Gregg Rothermel}, \bibinfo{person}{Mary Shaw}, {and} \bibinfo{person}{Susan Wiedenbeck}.} \bibinfo{year}{2011}\natexlab{}.
\newblock \showarticletitle{The state of the art in end-user software engineering}.
\newblock \bibinfo{journal}{\emph{ACM Computing Surveys (CSUR)}}  \bibinfo{volume}{43} (\bibinfo{year}{2011}), \bibinfo{pages}{1 -- 44}.
\newblock
\urldef\tempurl%
\url{https://api.semanticscholar.org/CorpusId:128364433}
\showURL{%
\tempurl}


\bibitem[Ko and Myers(2004)]%
        {ko2004designing}
\bibfield{author}{\bibinfo{person}{Amy~J Ko} {and} \bibinfo{person}{Brad~A Myers}.} \bibinfo{year}{2004}\natexlab{}.
\newblock \showarticletitle{Designing the whyline: a debugging interface for asking questions about program behavior}. In \bibinfo{booktitle}{\emph{Proceedings of the SIGCHI conference on Human factors in computing systems}}. \bibinfo{pages}{151--158}.
\newblock


\bibitem[Lee et~al\mbox{.}(2023)]%
        {lee2023deimos}
\bibfield{author}{\bibinfo{person}{Benjamin Lee}, \bibinfo{person}{Arvind Satyanarayan}, \bibinfo{person}{Maxime Cordeil}, \bibinfo{person}{Arnaud Prouzeau}, \bibinfo{person}{Bernhard Jenny}, {and} \bibinfo{person}{Tim Dwyer}.} \bibinfo{year}{2023}\natexlab{}.
\newblock \showarticletitle{Deimos: A grammar of dynamic embodied immersive visualisation morphs and transitions}. In \bibinfo{booktitle}{\emph{Proceedings of the 2023 CHI Conference on Human Factors in Computing Systems}}. \bibinfo{pages}{1--18}.
\newblock


\bibitem[Lehtinen et~al\mbox{.}(2024)]%
        {lehtinen2024let}
\bibfield{author}{\bibinfo{person}{Teemu Lehtinen}, \bibinfo{person}{Charles Koutcheme}, {and} \bibinfo{person}{Arto Hellas}.} \bibinfo{year}{2024}\natexlab{}.
\newblock \showarticletitle{Let's Ask AI About Their Programs: Exploring ChatGPT's Answers To Program Comprehension Questions}. In \bibinfo{booktitle}{\emph{Proceedings of the 46th International Conference on Software Engineering: Software Engineering Education and Training}}. \bibinfo{pages}{221--232}.
\newblock


\bibitem[Lewis(1992)]%
        {lewis1992psychometric}
\bibfield{author}{\bibinfo{person}{James~R Lewis}.} \bibinfo{year}{1992}\natexlab{}.
\newblock \showarticletitle{Psychometric evaluation of the post-study system usability questionnaire: The PSSUQ}. In \bibinfo{booktitle}{\emph{Proceedings of the human factors society annual meeting}}, Vol.~\bibinfo{volume}{36}. Sage Publications Sage CA: Los Angeles, CA, \bibinfo{pages}{1259--1260}.
\newblock


\bibitem[Liu et~al\mbox{.}(2023)]%
        {Liu2023WhatIW}
\bibfield{author}{\bibinfo{person}{Michael~Xieyang Liu}, \bibinfo{person}{Advait Sarkar}, \bibinfo{person}{Carina Negreanu}, \bibinfo{person}{Benjamin~G. Zorn}, \bibinfo{person}{J. Williams}, \bibinfo{person}{Neil Toronto}, {and} \bibinfo{person}{Andrew~D. Gordon}.} \bibinfo{year}{2023}\natexlab{}.
\newblock \showarticletitle{“What It Wants Me To Say”: Bridging the Abstraction Gap Between End-User Programmers and Code-Generating Large Language Models}.
\newblock \bibinfo{journal}{\emph{Proceedings of the 2023 CHI Conference on Human Factors in Computing Systems}} (\bibinfo{year}{2023}).
\newblock
\urldef\tempurl%
\url{https://api.semanticscholar.org/CorpusID:258107840}
\showURL{%
\tempurl}


\bibitem[Lyu et~al\mbox{.}(2023)]%
        {lyu2023faithful}
\bibfield{author}{\bibinfo{person}{Qing Lyu}, \bibinfo{person}{Shreya Havaldar}, \bibinfo{person}{Adam Stein}, \bibinfo{person}{Li Zhang}, \bibinfo{person}{Delip Rao}, \bibinfo{person}{Eric Wong}, \bibinfo{person}{Marianna Apidianaki}, {and} \bibinfo{person}{Chris Callison-Burch}.} \bibinfo{year}{2023}\natexlab{}.
\newblock \showarticletitle{Faithful chain-of-thought reasoning}. In \bibinfo{booktitle}{\emph{The 13th International Joint Conference on Natural Language Processing and the 3rd Conference of the Asia-Pacific Chapter of the Association for Computational Linguistics (IJCNLP-AACL 2023)}}.
\newblock


\bibitem[Maloney et~al\mbox{.}(2010)]%
        {maloney2010scratch}
\bibfield{author}{\bibinfo{person}{John Maloney}, \bibinfo{person}{Mitchel Resnick}, \bibinfo{person}{Natalie Rusk}, \bibinfo{person}{Brian Silverman}, {and} \bibinfo{person}{Evelyn Eastmond}.} \bibinfo{year}{2010}\natexlab{}.
\newblock \showarticletitle{The scratch programming language and environment}.
\newblock \bibinfo{journal}{\emph{ACM Transactions on Computing Education (TOCE)}} \bibinfo{volume}{10}, \bibinfo{number}{4} (\bibinfo{year}{2010}), \bibinfo{pages}{1--15}.
\newblock


\bibitem[Meta(2024)]%
        {react}
\bibfield{author}{\bibinfo{person}{Meta}.} \bibinfo{year}{2024}\natexlab{}.
\newblock \bibinfo{title}{React}.
\newblock \bibinfo{howpublished}{\url{https://react.dev/}}.
\newblock


\bibitem[MUI(2024)]%
        {mui}
\bibfield{author}{\bibinfo{person}{MUI}.} \bibinfo{year}{2024}\natexlab{}.
\newblock \bibinfo{title}{Material UI}.
\newblock \bibinfo{howpublished}{\url{https://mui.com/material-ui/}}.
\newblock


\bibitem[Myers et~al\mbox{.}(2006)]%
        {myers2006invited}
\bibfield{author}{\bibinfo{person}{Brad~A Myers}, \bibinfo{person}{Amy~J Ko}, {and} \bibinfo{person}{Margaret~M Burnett}.} \bibinfo{year}{2006}\natexlab{}.
\newblock \showarticletitle{Invited research overview: end-user programming}. In \bibinfo{booktitle}{\emph{CHI'06 extended abstracts on Human factors in computing systems}}. \bibinfo{pages}{75--80}.
\newblock


\bibitem[Ritschel et~al\mbox{.}(2022)]%
        {Ritschel2022CanGD}
\bibfield{author}{\bibinfo{person}{Nico Ritschel}, \bibinfo{person}{Felipe Fronchetti}, \bibinfo{person}{Reid Holmes}, \bibinfo{person}{Ronald Garcia}, {and} \bibinfo{person}{David~C. Shepherd}.} \bibinfo{year}{2022}\natexlab{}.
\newblock \showarticletitle{Can guided decomposition help end-users write larger block-based programs? a mobile robot experiment}.
\newblock \bibinfo{journal}{\emph{Proceedings of the ACM on Programming Languages}}  \bibinfo{volume}{6} (\bibinfo{year}{2022}), \bibinfo{pages}{233 -- 258}.
\newblock
\urldef\tempurl%
\url{https://api.semanticscholar.org/CorpusID:253239351}
\showURL{%
\tempurl}


\bibitem[Rosenberg et~al\mbox{.}(2024)]%
        {rosenberg2024drawtalking}
\bibfield{author}{\bibinfo{person}{Karl~Toby Rosenberg}, \bibinfo{person}{Rubaiat~Habib Kazi}, \bibinfo{person}{Li-Yi Wei}, \bibinfo{person}{Haijun Xia}, {and} \bibinfo{person}{Ken Perlin}.} \bibinfo{year}{2024}\natexlab{}.
\newblock \showarticletitle{DrawTalking: Building Interactive Worlds by Sketching and Speaking}.
\newblock \bibinfo{journal}{\emph{arXiv preprint arXiv:2401.05631}} (\bibinfo{year}{2024}).
\newblock


\bibitem[Ruiz(2021)]%
        {tldraw}
\bibfield{author}{\bibinfo{person}{Steve Ruiz}.} \bibinfo{year}{2021}\natexlab{}.
\newblock \bibinfo{title}{tldraw: A tiny little drawing app}.
\newblock \bibinfo{howpublished}{\url{https://tldraw.com/}}.
\newblock


\bibitem[Saquib et~al\mbox{.}(2021)]%
        {saquib2021constructing}
\bibfield{author}{\bibinfo{person}{Nazmus Saquib}, \bibinfo{person}{Rubaiat~Habib Kazi}, \bibinfo{person}{Li-yi Wei}, \bibinfo{person}{Gloria Mark}, {and} \bibinfo{person}{Deb Roy}.} \bibinfo{year}{2021}\natexlab{}.
\newblock \showarticletitle{Constructing embodied algebra by sketching}. In \bibinfo{booktitle}{\emph{Proceedings of the 2021 CHI Conference on Human Factors in Computing Systems}}. \bibinfo{pages}{1--16}.
\newblock


\bibitem[Sarkar et~al\mbox{.}(2022)]%
        {sarkar2022like}
\bibfield{author}{\bibinfo{person}{Advait Sarkar}, \bibinfo{person}{Andrew~D Gordon}, \bibinfo{person}{Carina Negreanu}, \bibinfo{person}{Christian Poelitz}, \bibinfo{person}{Sruti~Srinivasa Ragavan}, {and} \bibinfo{person}{Ben Zorn}.} \bibinfo{year}{2022}\natexlab{}.
\newblock \showarticletitle{What is it like to program with artificial intelligence?}
\newblock \bibinfo{journal}{\emph{arXiv preprint arXiv:2208.06213}} (\bibinfo{year}{2022}).
\newblock


\bibitem[Seffah et~al\mbox{.}(2005)]%
        {seffah2005human}
\bibfield{author}{\bibinfo{person}{Ahmed Seffah}, \bibinfo{person}{Jan Gulliksen}, {and} \bibinfo{person}{Michel~C Desmarais}.} \bibinfo{year}{2005}\natexlab{}.
\newblock \bibinfo{booktitle}{\emph{Human-centered software engineering-integrating usability in the software development lifecycle}}. Vol.~\bibinfo{volume}{8}.
\newblock \bibinfo{publisher}{Springer Science \& Business Media}.
\newblock


\bibitem[Smith et~al\mbox{.}(1994)]%
        {smith1994kidsim}
\bibfield{author}{\bibinfo{person}{David~Canfield Smith}, \bibinfo{person}{Allen Cypher}, {and} \bibinfo{person}{Jim Spohrer}.} \bibinfo{year}{1994}\natexlab{}.
\newblock \showarticletitle{KidSim: Programming agents without a programming language}.
\newblock \bibinfo{journal}{\emph{Commun. ACM}} \bibinfo{volume}{37}, \bibinfo{number}{7} (\bibinfo{year}{1994}), \bibinfo{pages}{54--67}.
\newblock


\bibitem[Song et~al\mbox{.}(2023)]%
        {song2023empirical}
\bibfield{author}{\bibinfo{person}{Da Song}, \bibinfo{person}{Zijie Zhou}, \bibinfo{person}{Zhijie Wang}, \bibinfo{person}{Yuheng Huang}, \bibinfo{person}{Shengmai Chen}, \bibinfo{person}{Bonan Kou}, \bibinfo{person}{Lei Ma}, {and} \bibinfo{person}{Tianyi Zhang}.} \bibinfo{year}{2023}\natexlab{}.
\newblock \showarticletitle{An empirical study of code generation errors made by large language models}. In \bibinfo{booktitle}{\emph{7th Annual Symposium on Machine Programming}}.
\newblock


\bibitem[Subramonyam et~al\mbox{.}(2024)]%
        {subramonyam2024bridging}
\bibfield{author}{\bibinfo{person}{Hari Subramonyam}, \bibinfo{person}{Roy Pea}, \bibinfo{person}{Christopher Pondoc}, \bibinfo{person}{Maneesh Agrawala}, {and} \bibinfo{person}{Colleen Seifert}.} \bibinfo{year}{2024}\natexlab{}.
\newblock \showarticletitle{Bridging the Gulf of Envisioning: Cognitive Challenges in Prompt Based Interactions with LLMs}. In \bibinfo{booktitle}{\emph{Proceedings of the CHI Conference on Human Factors in Computing Systems}}. \bibinfo{pages}{1--19}.
\newblock


\bibitem[Subramonyam et~al\mbox{.}(2020)]%
        {subramonyam2020texsketch}
\bibfield{author}{\bibinfo{person}{Hariharan Subramonyam}, \bibinfo{person}{Colleen Seifert}, \bibinfo{person}{Priti Shah}, {and} \bibinfo{person}{Eytan Adar}.} \bibinfo{year}{2020}\natexlab{}.
\newblock \showarticletitle{Texsketch: Active diagramming through pen-and-ink annotations}. In \bibinfo{booktitle}{\emph{Proceedings of the 2020 CHI Conference on Human Factors in Computing Systems}}. \bibinfo{pages}{1--13}.
\newblock


\bibitem[Suh et~al\mbox{.}(2022)]%
        {suh2022codetoon}
\bibfield{author}{\bibinfo{person}{Sangho Suh}, \bibinfo{person}{Jian Zhao}, {and} \bibinfo{person}{Edith Law}.} \bibinfo{year}{2022}\natexlab{}.
\newblock \showarticletitle{Codetoon: Story ideation, auto comic generation, and structure mapping for code-driven storytelling}. In \bibinfo{booktitle}{\emph{Proceedings of the 35th Annual ACM Symposium on User Interface Software and Technology}}. \bibinfo{pages}{1--16}.
\newblock


\bibitem[Suzuki et~al\mbox{.}(2020)]%
        {suzuki2020realitysketch}
\bibfield{author}{\bibinfo{person}{Ryo Suzuki}, \bibinfo{person}{Rubaiat~Habib Kazi}, \bibinfo{person}{Li-Yi Wei}, \bibinfo{person}{Stephen DiVerdi}, \bibinfo{person}{Wilmot Li}, {and} \bibinfo{person}{Daniel Leithinger}.} \bibinfo{year}{2020}\natexlab{}.
\newblock \showarticletitle{Realitysketch: Embedding responsive graphics and visualizations in AR through dynamic sketching}. In \bibinfo{booktitle}{\emph{Proceedings of the 33rd Annual ACM Symposium on User Interface Software and Technology}}. \bibinfo{pages}{166--181}.
\newblock


\bibitem[Sveidqvist and contributors(2014)]%
        {mermaidjs}
\bibfield{author}{\bibinfo{person}{Knut Sveidqvist} {and} \bibinfo{person}{Mermaid.js contributors}.} \bibinfo{year}{2014}\natexlab{}.
\newblock \bibinfo{title}{Mermaid: Generation of diagrams and flowcharts from text in a similar manner as markdown}.
\newblock \bibinfo{howpublished}{\url{https://mermaid.js.org/}}.
\newblock


\bibitem[Tan et~al\mbox{.}(2023)]%
        {tan2023copilot}
\bibfield{author}{\bibinfo{person}{Chee~Wei Tan}, \bibinfo{person}{Shangxin Guo}, \bibinfo{person}{Man~Fai Wong}, {and} \bibinfo{person}{Ching~Nam Hang}.} \bibinfo{year}{2023}\natexlab{}.
\newblock \showarticletitle{Copilot for Xcode: exploring AI-assisted programming by prompting cloud-based large language models}.
\newblock \bibinfo{journal}{\emph{arXiv preprint arXiv:2307.14349}} (\bibinfo{year}{2023}).
\newblock


\bibitem[Tan and Subramonyam(2024)]%
        {tan2024more}
\bibfield{author}{\bibinfo{person}{Mei Tan} {and} \bibinfo{person}{Hari Subramonyam}.} \bibinfo{year}{2024}\natexlab{}.
\newblock \showarticletitle{More than model documentation: uncovering teachers' bespoke information needs for informed classroom integration of ChatGPT}. In \bibinfo{booktitle}{\emph{Proceedings of the CHI Conference on Human Factors in Computing Systems}}. \bibinfo{pages}{1--19}.
\newblock


\bibitem[Tian et~al\mbox{.}(2024)]%
        {tian2024debugbench}
\bibfield{author}{\bibinfo{person}{Runchu Tian}, \bibinfo{person}{Yining Ye}, \bibinfo{person}{Yujia Qin}, \bibinfo{person}{Xin Cong}, \bibinfo{person}{Yankai Lin}, \bibinfo{person}{Yinxu Pan}, \bibinfo{person}{Yesai Wu}, \bibinfo{person}{Haotian Hui}, \bibinfo{person}{Weichuan Liu}, \bibinfo{person}{Zhiyuan Liu}, {et~al\mbox{.}}} \bibinfo{year}{2024}\natexlab{}.
\newblock \showarticletitle{Debugbench: Evaluating debugging capability of large language models}.
\newblock \bibinfo{journal}{\emph{arXiv preprint arXiv:2401.04621}} (\bibinfo{year}{2024}).
\newblock


\bibitem[Vrettakis et~al\mbox{.}(2020)]%
        {vrettakis2020story}
\bibfield{author}{\bibinfo{person}{Ektor Vrettakis}, \bibinfo{person}{Christos Lougiakis}, \bibinfo{person}{Akrivi Katifori}, \bibinfo{person}{Vassilis Kourtis}, \bibinfo{person}{Stamatis Christoforidis}, \bibinfo{person}{Manos Karvounis}, {and} \bibinfo{person}{Yannis Ioanidis}.} \bibinfo{year}{2020}\natexlab{}.
\newblock \showarticletitle{The story maker-an authoring tool for multimedia-rich interactive narratives}. In \bibinfo{booktitle}{\emph{Interactive Storytelling: 13th International Conference on Interactive Digital Storytelling, ICIDS 2020, Bournemouth, UK, November 3--6, 2020, Proceedings 13}}. Springer, \bibinfo{pages}{349--352}.
\newblock


\bibitem[Wei et~al\mbox{.}(2022)]%
        {wei2022chain}
\bibfield{author}{\bibinfo{person}{Jason Wei}, \bibinfo{person}{Xuezhi Wang}, \bibinfo{person}{Dale Schuurmans}, \bibinfo{person}{Maarten Bosma}, \bibinfo{person}{Fei Xia}, \bibinfo{person}{Ed Chi}, \bibinfo{person}{Quoc~V Le}, \bibinfo{person}{Denny Zhou}, {et~al\mbox{.}}} \bibinfo{year}{2022}\natexlab{}.
\newblock \showarticletitle{Chain-of-thought prompting elicits reasoning in large language models}.
\newblock \bibinfo{journal}{\emph{Advances in neural information processing systems}}  \bibinfo{volume}{35} (\bibinfo{year}{2022}), \bibinfo{pages}{24824--24837}.
\newblock


\bibitem[Weyssow et~al\mbox{.}(2022)]%
        {Weyssow2022BetterMT}
\bibfield{author}{\bibinfo{person}{Martin Weyssow}, \bibinfo{person}{Houari~A. Sahraoui}, {and} \bibinfo{person}{Bang Liu}.} \bibinfo{year}{2022}\natexlab{}.
\newblock \showarticletitle{Better Modeling the Programming World with Code Concept Graphs-augmented Multi-modal Learning}.
\newblock \bibinfo{journal}{\emph{2022 IEEE/ACM 44th International Conference on Software Engineering: New Ideas and Emerging Results (ICSE-NIER)}} (\bibinfo{year}{2022}), \bibinfo{pages}{21--25}.
\newblock
\urldef\tempurl%
\url{https://api.semanticscholar.org/CorpusID:245837887}
\showURL{%
\tempurl}


\bibitem[White et~al\mbox{.}(2023)]%
        {white2023prompt}
\bibfield{author}{\bibinfo{person}{Jules White}, \bibinfo{person}{Quchen Fu}, \bibinfo{person}{Sam Hays}, \bibinfo{person}{Michael Sandborn}, \bibinfo{person}{Carlos Olea}, \bibinfo{person}{Henry Gilbert}, \bibinfo{person}{Ashraf Elnashar}, \bibinfo{person}{Jesse Spencer-Smith}, {and} \bibinfo{person}{Douglas~C Schmidt}.} \bibinfo{year}{2023}\natexlab{}.
\newblock \showarticletitle{A prompt pattern catalog to enhance prompt engineering with chatgpt}.
\newblock \bibinfo{journal}{\emph{arXiv preprint arXiv:2302.11382}} (\bibinfo{year}{2023}).
\newblock


\bibitem[White et~al\mbox{.}(2024)]%
        {white2024chatgpt}
\bibfield{author}{\bibinfo{person}{Jules White}, \bibinfo{person}{Sam Hays}, \bibinfo{person}{Quchen Fu}, \bibinfo{person}{Jesse Spencer-Smith}, {and} \bibinfo{person}{Douglas~C Schmidt}.} \bibinfo{year}{2024}\natexlab{}.
\newblock \showarticletitle{Chatgpt prompt patterns for improving code quality, refactoring, requirements elicitation, and software design}.
\newblock In \bibinfo{booktitle}{\emph{Generative ai for effective software development}}. \bibinfo{publisher}{Springer}, \bibinfo{pages}{71--108}.
\newblock


\bibitem[Wongsuphasawat et~al\mbox{.}(2015)]%
        {wongsuphasawat2015voyager}
\bibfield{author}{\bibinfo{person}{Kanit Wongsuphasawat}, \bibinfo{person}{Dominik Moritz}, \bibinfo{person}{Anushka Anand}, \bibinfo{person}{Jock Mackinlay}, \bibinfo{person}{Bill Howe}, {and} \bibinfo{person}{Jeffrey Heer}.} \bibinfo{year}{2015}\natexlab{}.
\newblock \showarticletitle{Voyager: Exploratory analysis via faceted browsing of visualization recommendations}.
\newblock \bibinfo{journal}{\emph{IEEE transactions on visualization and computer graphics}} \bibinfo{volume}{22}, \bibinfo{number}{1} (\bibinfo{year}{2015}), \bibinfo{pages}{649--658}.
\newblock


\bibitem[Xu et~al\mbox{.}(2024)]%
        {xu2024hallucination}
\bibfield{author}{\bibinfo{person}{Ziwei Xu}, \bibinfo{person}{Sanjay Jain}, {and} \bibinfo{person}{Mohan Kankanhalli}.} \bibinfo{year}{2024}\natexlab{}.
\newblock \showarticletitle{Hallucination is inevitable: An innate limitation of large language models}.
\newblock \bibinfo{journal}{\emph{arXiv preprint arXiv:2401.11817}} (\bibinfo{year}{2024}).
\newblock


\bibitem[Yin et~al\mbox{.}(2024)]%
        {yin2024rectifier}
\bibfield{author}{\bibinfo{person}{Xin Yin}, \bibinfo{person}{Chao Ni}, \bibinfo{person}{Tien~N Nguyen}, \bibinfo{person}{Shaohua Wang}, {and} \bibinfo{person}{Xiaohu Yang}.} \bibinfo{year}{2024}\natexlab{}.
\newblock \showarticletitle{Rectifier: Code translation with corrector via llms}.
\newblock \bibinfo{journal}{\emph{arXiv preprint arXiv:2407.07472}} (\bibinfo{year}{2024}).
\newblock


\bibitem[Zhu et~al\mbox{.}(2020)]%
        {Zhu2020HierarchicalPF}
\bibfield{author}{\bibinfo{person}{Yifeng Zhu}, \bibinfo{person}{Jonathan Tremblay}, \bibinfo{person}{Stan Birchfield}, {and} \bibinfo{person}{Yuke Zhu}.} \bibinfo{year}{2020}\natexlab{}.
\newblock \showarticletitle{Hierarchical Planning for Long-Horizon Manipulation with Geometric and Symbolic Scene Graphs}.
\newblock \bibinfo{journal}{\emph{2021 IEEE International Conference on Robotics and Automation (ICRA)}} (\bibinfo{year}{2020}), \bibinfo{pages}{6541--6548}.
\newblock
\urldef\tempurl%
\url{https://api.semanticscholar.org/CorpusID:229156165}
\showURL{%
\tempurl}


\end{thebibliography}
